\documentclass[iop,12pt]{emulateapj}
\usepackage{apjfonts}
\usepackage{natbib}
\usepackage{amsfonts}
\usepackage{amsmath}
\usepackage{multirow}
\usepackage{enumerate}
\usepackage[usenames]{color}
\def\Dwa{$\,$\uppercase\expandafter{\romannumeral5}$\,$}

\def\sless{\lower2pt\hbox{$\buildrel {\scriptstyle <}
   \over {\scriptstyle\sim}$}}

\def\sgreat{\lower2pt\hbox{$\buildrel {\scriptstyle >}
   \over {\scriptstyle\sim}$}}
\def\sharpnull#1{}

\def\sref{\S\ref}
\def\fref{Fig.~\ref}

\def\tref{Table~\ref}
\def\esref{Eqs.~\ref}
\def\eref{Eq.~\ref}
\newcommand\cqg{\rm{Class. Quantum Grav.}}%
\newcommand{\code}[1]{\texttt{#1}}

\begin{document}
%-----------------------------------------------------
\slugcomment{Accepted to ApJS, 2015}

\title{An Open-Source Neutrino Radiation Hydrodynamics Code for
  Core-Collapse Supernovae}

\author{Evan O'Connor\altaffilmark{1,2,3}}
 \altaffiltext{1}{North Carolina State University, Department of
   Physics, Campus Code 8202, Raleigh, North 
   Carolina, USA, 27695, evanoconnor@ncsu.edu}
 \altaffiltext{2}{CITA, Canadian 
    Institute for Theoretical Astrophysics, Toronto, Canada, M5S 3H8}
 \altaffiltext{3}{Hubble Fellow}

\begin{abstract}
  We present an open-source update to the spherically-symmetric,
  general-relativistic hydrodynamics, core-collapse supernova (CCSN)
  code \code{GR1D} (O'Connor \& Ott, 2010, CQG, 27, 114103). The
  source code is available at \url{http://www.GR1Dcode.org}. We extend
  its capabilities to include a general relativistic treatment of
  neutrino transport based on the moment formalisms of Shibata et al.,
  2011, PTP, 125, 1255 and Cardall et al., 2013, PRD, 87 103004. We
  pay special attention to implementing and testing numerical methods
  and approximations that lessen the computational demand of the
  transport scheme by removing the need to invert large matrices. This
  is especially important for the implementation and development of
  moment-like transport methods in two and three dimensions. A
  critical component of neutrino transport calculations are the
  neutrino-matter interaction coefficients that describe the
  production, absorption, scattering, and annihilation of neutrinos.
  In this article we also describe our open-source, neutrino
  interaction library \code{NuLib} (available at
  \url{http://www.nulib.org}). We believe that an open-source approach
  to describing these interactions is one of the major steps needed to
  progress towards robust models of CCSNe and robust predictions of
  the neutrino signal. We show, via comparisons to full Boltzmann
  neutrino transport simulations of CCSNe, that our neutrino transport
  code performs remarkably well. Furthermore, we show that the methods
  and approximations we employ to increase efficiency do not decrease
  the fidelity of our results. We also test the ability of our general
  relativistic transport code to model failed CCSNe by evolving a 40
  solar-mass progenitor to the onset of collapse to a black hole.
\end{abstract}


\section{Introduction}
\label{section:intro}

For most massive stars with a zero-age main sequence (ZAMS) mass
larger than 8--10\,$M_\odot$, the end of hydrostatic stellar evolution
is marked by one of the most energetic events in the modern universe,
a core-collapse supernova (CCSN). The cores of these massive stars
become unstable to gravitational collapse when the gravitational force
can no longer be balanced by the electron degeneracy pressure supplied
by the electrons in the inert iron core. This marks the beginning of
the core collapse phase of a CCSN. At this time, neutrinos begin to
shift from being a mere sink of leptons and energy during stellar
evolution to playing a much more dominant and complex role in setting
the structure, dynamics, and thermodynamics of the CCSN central engine
\citep{bethe:90}. Over 50 years of theoretical research on CCSN has
revealed that an accurate treatment of neutrinos is an important and
essential ingredient in modeling the CCSN central engine
\citep{bethewilson:85, mezzacappa:93b, liebendoerfer:04, janka:07,
  ott:08, mueller:10} and the protoneutron star cooling phase
\citep{fischer:10,huedepohl:10,roberts:12b}. Furthermore, a precision
treatment of neutrinos in CCSN models is needed to aid in the
interpretation of neutrino signals from galactic CCSNe and the diffuse
supernova neutrino background.

Energy loss due to neutrino emission, along side nuclear dissociation,
plays a large role in causing the supernova shock to initially
stall. However, neutrinos are also likely a key ingredient in the
ultimate shock revival. While the complete picture is still unclear,
the currently most favorable scenario for this revival involves the
neutrino mechanism \citep{bethewilson:85}. The essence of the neutrino
mechanism is the charged-current heating that electron-type neutrinos
and antineutrinos emitted from deep cooling regions impose on the
material behind the shock front. This heating increases the thermal
energy and pressure of the matter behind the shock and it also drives
convection and turbulence--all of these effects of neutrino heating
enable shock expansion and aid in the transition from the initial
accretion phase to the ultimate explosion phase in successful
CCSNe. We refer the reader to recent reviews for full details
\citep{janka:12a,burrows:13a}.

CCSNe and other high-energy astrophysical events like neutron
star-neutron star or neutron star-black hole mergers are environments
where one cannot treat neutrinos like one typically treats other
particles. For example, in the central engine of CCSNe, photons,
electrons, and nucleons are always in local thermodynamic
equilibrium. This allows us to assume the particle distribution
function and derive an equation of state (EOS) (eg. pressure, internal
energy, chemical potentials, ...). Since neutrinos thermodynamically
decouple from the matter at densities we are interested in studying,
we cannot assume thermodynamic equilibrium as we do for other
particles. In fact, it is precisely this aspect of neutrinos which
gives rise to the neutrino mechanism. Neutrinos decouple from matter
at one density and temperature and then non-locally transfer energy
and lepton number to another region. We must model the neutrino
distribution function itself to accurately simulate these
astrophysical events and capture all of the essential physics.

In order to properly deal with massless neutrinos one must solve for
the 7-dimensional neutrino distribution function
($f_{\nu_i}(x^\mu,p^\alpha)$) as a function of time ($x^0 = t$) for
all spatial points ($x^j = \vec{x}$), for all neutrino energies and
propagation angles ($p^\alpha = (E_{\nu_i},\vec{p})$; with the
restriction that $E_{\nu_i} = |\vec{p}|$), and for each species
($\nu_i$). The time evolution of $f_{\nu_i}(x^\mu,p^\alpha)$ is
governed by the relativistic collisional Boltzmann transport equation
\citep{lindquist:66},
\begin{equation}
  p^\alpha \left[\frac{\partial f_{\nu_i}}{\partial x^\alpha} -
  \Gamma^{\beta}_{\ \,\alpha \gamma}\ p^\gamma
  \frac{\partial f_\nu}{\partial p^\beta}\right] =
\left[\frac{df_\nu}{d\tau}\right]_\text{coll}\,,
\label{eq:CBTE}
\end{equation}
where $\Gamma^{...}$ are the Christoffel symbols and external
influences including the scattering of neutrinos with other neutrinos
or the surrounding matter and the absorption and emission of neutrinos
by the matter (i.e., collisional processes) are apart of the
collisional term, $[df_\nu/d\tau]_\mathrm{coll}$.

In practice, solving the full time evolution of the Boltzmann equation
for neutrinos in the CCSN context is a formidable task. For typical
post-bounce configurations and neutrino energies, the neutrino
distribution function transitions from its thermal equilibrium and
isotropic value in the protoneutron star core ($r \lesssim 30$\,km),
to essentially decoupled (optical depth $\tau \sim 0.1$) in the gain region
($r \sim 100$\,km) and forward peaked at distances of $\gtrsim$200\,km
\citep{thompson:03,sumiyoshi:12}. The dramatic change in the behavior
of the equations from being dominated by the collision terms to being
dominated by the transport terms over a short distance make the
numerics particularly difficult as many approximations typically used
for radiation (like the diffusion limit) are not valid everywhere. As
mentioned above, the transition from the diffusive regime to the free
streaming regime is absolutely crucial for the neutrino mechanism of
CCSNe. It is in this region where interactions between the neutrino
field and the matter are still appreciable and can lead to the
development of a heating region where a net positive amount of energy
can be transferred from the neutrino field to the matter. It is
commonplace to make approximations to simplify the calculation from
the fully relativistic 3+2+1+1 (3 spatial dimensions, 2 neutrino
propagation angles, neutrino energy, and the time dimension) problem
for each neutrino species down to a more tractable problem. We briefly
discuss the various neutrino transport schemes used in the CCSN
supernova community and discuss the advantages and disadvantages of
each. We start with the most approximate and then increase in
complexity of the scheme.

{\emph{Leakage Schemes:}} Perhaps the crudest approximation to
neutrino transport, neutrino leakage, cannot really be called a
transport method at all as no evolution of the neutrino distribution
function actually occurs. In general, a leakage scheme estimates the
local neutrino energy and number emission rates by an interpolation
between the free emission rate and the emission rate based on the
diffusion approximation. This emitted energy and lepton number is then
explicitly extracted from the matter. The leakage scheme used in
\code{GR1D} is described in detail in \cite{oconnor:10}. Various other
examples of leakage schemes used in the high-energy astrophysics
community include \cite{ruffert:96,rosswog:03b,sekiguchi:10} A
disadvantage of neutrino leakage schemes in their purest form is that
they cannot self-consistently reproduce the neutrino heating. In
spherically symmetric problems, or in problems that are largely
spherical (like core collapse in multiple dimensions) this can be
overcome by integrating the luminosity coming from smaller radii, as
we have done with the leakage scheme in \code{GR1D}. However, in
simulations with much less symmetry in the matter distributions (like
accretion disks or compact-object mergers) this is not possible
without resorting to methods like ray tracing \citep{perego:14}.

{\emph{Moment Schemes:}} An approximation often made in neutrino
transport is to remove the full angular dependence of the Boltzmann
transport equation by expanding the neutrino distribution function as
a series of moments. The zeroth moment in this expansion is,
\begin{equation} 
\mathcal{J}_{\nu}(\vec{x}, \epsilon_\nu, t)=\frac{\epsilon_{\nu}^3}{4\pi (hc)^3}\int_\Omega
f_{\nu} (\vec{x}, \vec{p}, t)
d\Omega\,,
\end{equation}
which is a scalar quantity (the spectral energy density) that depends
only on the spatial location ($\vec{x}$), neutrino energy
($|\vec{p}| = \epsilon_\nu$), and time ($t$). This reduces the degrees
of freedom from 7 to 5 (in three dimensions; 4 to 3 in spherical
symmetry). This moment expansion can continue to higher moments, for
example the next moment is a vector (the spectral momentum density)
that depends only on the spatial location, neutrino energy, and
time. Moment methods are convenient as they reduce the dimensionality
of the problem resulting in fewer equations to solve. Also, the first few moments
have an intuitive physical meaning and capture much of the physics in
many astrophysical problems.

Within the moment scheme framework there is a lot of room for further
approximations. For example, one can truncate the moment expansion at
any order by specifying a \emph{closure}, an expression that
approximates the $n+1$ moment as a function of the first $n$ moments.
The simplest variant, where one closes the moment expansion after the
zeroth moment, is flux-limited diffusion (FLD). Examples in the CCSN
context include \cite{bruenn:85} in 1D and
\cite{fryer:99,burrows:07a,swesty:09,yakunin:10,zhang:13} in 2D. In
FLD schemes, the underlying equation one solves is a diffusion
equation for the neutrino energy density. A flux-limiter must be
invoked that ensures the radiation does not travel faster than the
speed of light in regions where the diffusion approximation fails. 
Similar schemes to FLD are the work of \cite{scheck:06} and \cite{mueller:15}.
In the former, the
transport equation for the zeroth moment is solved in the low optical
depth limit and a boundary condition is imposed at high optical
depth. In the latter, an equation for the energy-dependent zeroth moment of
the neutrino distribution function is solved with the help of a
closure for determining the flux factor.
 Another FLD-like method is the isotropic diffusion source
approximation (IDSA) \citep{liebendoerfer:09} which evolves two
components of the neutrino distribution function, a trapped component
and a free streaming component and transfers neutrinos between these
components via empirical source terms. The
M1 moment scheme for neutrino radiation transport
\citep{pons:00,obergaulinger:14,kuroda:12,oconnor:13,just:15,kuroda:15}, evolves
both the energy density and the momentum density but assumes an
analytic closure for required higher moments. One can also define
moment schemes where the closure is not analytic, but rather one has a
variable Eddington tensor
\citep{burrows:00,thompson:03,rampp:02,buras:06a,buras:06b}. In these
cases, one uses the evolved moments as source terms to a formal
integration of a Boltzmann-like equation. From this solution the
higher moments (and therefore the Eddington tensor) are calculated and
the system is iterated until convergence is reached. Depending on the
methods used to solve for the closure, this method can be equivalent
to a full Boltzmann neutrino transport calculation. 

{\emph{Boltzmann Schemes:}} It is also possible to solve the full
Boltzmann equation taking explicitly into account both the energy and
angular dependence of the neutrino distribution function. This has
been done in spherical symmetry \citep{mezzacappa:93a, mezzacappa:93b,
  mezzacappa:93c, yamada:97, liebendoerfer:04, sumiyoshi:05}, in 2D
\citep{livne:04,ott:08,brandt:11}, and recently in 3D
\citep{sumiyoshi:12,sumiyoshi:15}. The 2D and 3D angle dependent works
ignore velocity terms and do not couple the energy groups. The 3D work of
\citet{sumiyoshi:12,sumiyoshi:15} is a challenging task and present
results are only for static backgrounds with very low resolution in
all quantities considered. The high dimensionality of Boltzmann
schemes make them very expensive and simulating a three dimensional CCSN
central engine with adequate resolution using this treatment is
prohibitive in the foreseeable future.

{\emph{Monte Carlo Schemes:}} Like many numerical problems, the
transport of neutrinos can be solved by throwing computational power at
the problem. For this, Monte Carlo methods for solving the Boltzmann
transport equation can be used. Monte Carlo scales almost perfectly
to large problem sizes and therefore is very attractive for 3D
simulations. Monte Carlo methods have been used in the CCSN context
for many years \citep{janka:89c, janka:89d, janka:92c,
  abdikamalov:12}  are potentially promising for large-scale 3D
simulations in the future, although there is currently no published
Monte Carlo neutrino transport work in multiple dimensions.

We note that, in principle, each of these schemes can be either energy
dependent or energy independent. The latter are referred to as
\emph{grey} transport or \emph{grey} leakage. Grey methods are
computationally appealing as they remove an entire dimension of the
problem.  However, the energy dependence of the neutrino transport problem in
the early phase of core-collapse supernovae is crucial. The cross
sections of neutrinos with matter vary strongly with energy ($\propto
\epsilon^2$). This, in combination with the matter profiles in an
accreting protoneutron star, leads to a spatial separation of the
neutrinosphere locations and results in neutrino spectra that are
quite non-thermal. Grey transport schemes cannot model this and
therefore overestimate the number of high energy neutrinos in the gain
region. In fact, early grey transport schemes in 2D were successful in
obtaining explosions \citep{herant:94, bhf:95,
  fryer:99} while energy dependent simulations were much less
energetic \citep{buras:06a}. This aspect of neutrino transport in core
collapse warrants a multi-energy (or often referred to as multi-group)
treatment of neutrinos. 
One of the largest deficiencies in the previous version of \code{GR1D}
was its treatment of neutrinos. While the leakage/heating scheme is
efficient and captures qualitative and globally quantitative aspects
of the post-bounce phase, it cannot be trusted and it does not
predict, for example, the emitted energy spectrum of neutrinos. In
this paper, we present an implementation of a fully general
relativistic, energy dependent, multi-species neutrino radiation
transport code built on the general relativistic hydrodynamics code
\code{GR1D}. For each neutrino species and energy group, we evolve the
first two moments of the neutrino distribution function and close the
system of resulting equations via an analytic closure. We pay special
attention to developing explicit methods or approximations, where
possible, that will enable efficient scaling to higher dimensions. We
compare our code, as well as our approximation methods, to full
Boltzmann neutrino transport simulations of core collapse and several
standard radiation test cases. We achieve remarkable agreement in both
hydrodynamic and radiation quantities. Our complete code is available
online at \url{http://www.gr1dcode.org}. We also make our neutrino
interaction library \code{NuLib} open-source. It is available at
\url{http://www.nulib.org}. Details of both codes pertaining to this
particular paper, including the parameter files needed to reproduce
all of the results of this paper are available at
\url{http://www.stellarcollapse.org/GR1Dv2}.  We encourage anyone who
uses these codes to be as forthcoming with their results as we are by
also providing their parameter files and any changes to the code
needed to exactly reproduce their results.

In \sref{sec:methods} ,we briefly summarize the hydrodynamic code
\code{GR1D} and then go on to discuss in detail our new neutrino
transport methods. In \sref{sec:nulib}, we describe \code{NuLib}--an
open-source library for neutrino interaction rates. \sref{sec:tests}
and \sref{sec:ccsntests} is where we test our code against several
common radiation test problems and against full Boltzmann radiation
transport simulations of core collapse. We also show the results of
the evolution of a failed CCSN. We conclude in \sref{sec:conclusions}.

\section{Methods}
\label{sec:methods}
\subsection{\code{GR1D}'s Hydrodynamics}
\label{sec:GRhydro}
Before describing our neutrino transport scheme we briefly summarize
the general relativistic hydrodynamics code \code{GR1D}
\citep{oconnor:10}.  \code{GR1D} is based on the work of
\cite{romero:96} and uses the radial gauge, polar slicing metric
$g_{\alpha \beta} = \mathrm{diag}(-\alpha^2, X^2, r^2,
r^2\sin^2\theta)$.
In spherical symmetry, the metric coefficients can be determined
directly from Hamiltonian and momentum constraints and do not have to
be evolved.  The $g_{rr}$ component of the metric,
$X = 1/[1-2m(r)/r]^{1/2}$ is solved from an ordinary differential
equation for the enclosed gravitational mass,
\begin{equation}
\frac{dm}{dr} = 4\pi (\rho h W^2 - P + \tau_m^\nu)\,;\ \ \ \ m(0) = 0\,,\label{eq:mass}
\end{equation}
where $\rho$, $h$, $W$, and $P$ and the matter density, enthalpy,
Lorentz factor, and pressure, respectively.  The enthalpy,
$h=1+\epsilon+P/\rho$ contains the internal energy ($\epsilon$) of the
system. The Lorentz factor is related to the velocity of the fluid,
$W=1/[1-v^2]^{1/2}$ where following the convention of \code{GR1D} we
define the fluid velocity as $v=X v^r$.  Both the internal energy and
the pressure receive contributions from the nuclear EOS, electrons,
positrons, and photons. After solving for the enclosed gravitational mass,
the other metric factor, the lapse ($\alpha=\exp{(\phi)}$), can be
solved for via the momentum constraint,
\begin{equation}
  \frac{d\phi}{dr} = X^2\left[\frac{m}{r^2}+ 4\pi r
  (\rho h W^2 v^2 +P + \tau_\phi^\nu)\right]\,;\ \ \ \ \phi(\infty) = 0\,,\label{eq:phi}
\end{equation}
If we take $\tau^\nu_{m,\phi} = 0$, the expressions given here are for
a matter only stress energy tensor. We include the contribution from
neutrinos by solving Einstein's equation treating both the matter and
the neutrinos as sources $G^{\mu \nu} = 8\pi (T^{\mu
\nu}_\mathrm{matter} + T^{\mu \nu}_\mathrm{neutrinos})$. We delay
giving an expression for $T^{\mu \nu}_\mathrm{neutrinos}$, and
therefore $\tau^\nu_{m,\phi}$ until \sref{sec:numattercoupling} when
we are more acquainted with the moment formalism.

The hydrodynamic evolution equations are solved via the
flux-conservative Valencia formulation of relativistic hydrodynamics
\citep{font:00}. The full derivation of the evolution equations used in
\code{GR1D} is presented in Appendix A of \cite{oconnor:10}, we merely
present the equations here. The collection of evolution equations for
the hydrodynamic variables in \code{GR1D} are,
\begin{equation}
\partial_t \vec{\mathcal{U}} + \frac{1}{r^2}\partial_r\left[\frac{\alpha r^2
}{X}\vec{\mathcal{F}}(\vec{\mathcal{U}})\right] = \vec{\mathcal{S}}\,.
\end{equation}
Using $D= \rho W X$, $S^r = \rho h W^2 v$, and $\tau = \rho h W^2 - P
-D$, the state, flux, and source vectors are defined as,
\begin{eqnarray}
\label{eq:chv}
\vec{\mathcal{U}}&=&\{D,DY_e, S^r, \tau\}\,,\\ 
\vec{\mathcal{F}}(\vec{\mathcal{U}}) &=& \{Dv, DY_e v, S^r v + P,
S^r-Dv\}\,,\\
\nonumber
\vec{\mathcal{S}} &=& \{0,0,(S^r v -\tau - D)\alpha X
(8\pi r P + m/r^2) \\
&&\hspace*{1.6cm}+ \alpha P X m/r^2 + 2\alpha
  P/(X r),0\}\,.
\end{eqnarray}

For the simulations in this paper, in \code{GR1D}, we first perform a
hydrodynamic update using a second-order Runge-Kutta scheme as
described in \cite{oconnor:10}. For each substep of the Runge-Kutta,
we first reconstruct $\rho$, $Y_e$, and $\epsilon$ to the cell edges
using either the piecewise linear, total-variation-diminishing (TVD;
\citep{vanleer:77}) minmod method (during the collapse phase) or the
piecewise parabolic method (PPM \citealt{colella:84}; after the central
density reaches $10^{12}$\,g\,cm$^{-3}$).  The intercell fluxes are
determined by using these reconstructed values and the HLLE solution
to the Riemann problem \cite{HLLE:88}. 
After the hydrodynamic update to the $(n+1)$ time step is complete, we
update the neutrino radiation fields from the $(n)$ time step to the
$(n+1)$ time step using the methods described in the following
section.

\subsection{\code{GR1D}'s Neutrino Transport Scheme}

\code{GR1D}'s neutrino transport scheme follows from the formalisms
presented in \cite{shibata:11} and \cite{cardall:12rad}. These
formalisms differ in their derivation, however the resulting evolution
equations are identical. We will rely on both formalisms in our
implementation for their complementary and thorough derivations. For
example, only \cite{shibata:11} present a general casting of the
neutrino source terms in terms of neutrino distribution moments while
\cite{cardall:12rad} fully derive the energy-coupling terms for the
evolution equations. We immediately specialize the evolution equations
to the metric and notation of \code{GR1D}. We forgo use of a general
metric to avoid a mere restating of the derivations of
\cite{shibata:11} and \cite{cardall:12rad}. We hope that the
simplicity of \code{GR1D}'s metric, and our assumption of spherical
symmetry will make the system of evolution equations clear, but
general enough so that the numerical methods can easily be adapted to
hydrodynamic schemes and metrics other than those \code{GR1D} and that
they can be easily extended to multiple dimensions in the future.

We evolve the first two moments of the neutrino distribution function.
We do this for each spatial grid point, each neutrino species, and
each energy group.  We express the evolution equations in the
coordinate frame of \code{GR1D} while choosing the neutrino energy
spectrum coordinates (neutrino energies) in the comoving frame.  The
latter allows the neutrino interaction terms to be determined for
fixed energies regardless of the fluid velocity and then projected
into the coordinate frame. We close the system of moment equations
using an analytic closure. In the remainder of this section, we will
present the evolution equations and discuss the numerical techniques
implemented in \code{GR1D} to solve \emph{(a)} the closure relation,
\emph{(b)} the energy-group flux term, \emph{(c)} the spatial flux
term, and \emph{(d)} the geometric and neutrino source terms.

\subsubsection{Radiation Moment Evolution Equations}

We start by reexpressing Eqs.~171 and 172 of \cite{cardall:12rad} in
the notation of \code{GR1D}. This set of equations describes the
evolution of the energy-dependent zeroth and first lab-frame radiation
moments, the neutrino energy density ($E$) and the neutrino momentum
density ($F_r$), respectively. We forgo a complete rederivation of
this equation and point the reader to either \cite{cardall:12rad} or
\cite{shibata:11}.
\begin{eqnarray}
  \partial_t [E] + \frac{1}{r^2}\partial_r [\frac{\alpha r^2}{X^2}F_r]  +
  \partial_\epsilon [\epsilon (R_t + O_t)] &=&  G_t +
  C_t\,,\label{eq:e}\\
  \partial_t [F_r] + \frac{1}{r^2}\partial_r [\frac{\alpha r^2}{X^2} P_{rr}]  +
  \partial_\epsilon [\epsilon (R_r + O_r)] &=&  G_r + C_r\,.\label{eq:fr}
\end{eqnarray}
$R_\alpha$ and $O_\alpha$ are terms that originate from gravitational
redshifting and observer motions, respectively. $G_\alpha$ and
$C_\alpha$ describe source terms due to geometric and matter
interactions, respectively. $P_{rr}$ is the next highest moment of the
neutrino distribution function. In general the evolution equation for
a given moment will always depend on higher order moments. We describe
the method for calculating $P_{rr}$, or `closing the system of
equations' in \sref{sec:closure}. We explicitly note some subtle
differences between the notation of \cite{cardall:12rad} and our work.
We evolve the energy and momentum density of a particular energy group
directly, i.e. the cgs unit of $E$ and $F_r$ are
erg\,cm$^{-3}$\,sr$^{-1}$\,MeV$^{-1}$, and
erg\,cm$^{-2}$\,s$^{-1}$\,sr$^{-1}$\,MeV$^{-1}$, respectively.
Additionally, following the convention of \code{GR1D}, we evolve $E$
and $F_r$ rather then the densitized version $\sqrt{\gamma} E$ and
$\sqrt{\gamma} F_r$. This changes the geometric source terms $G_t$ and
$G_r$. We present our version of these expressions in
\sref{sec:geosourceterms}.

For the time evolution of the neutrino moments we use a simple
first-order implicit-explicit method. The spatial flux terms
$\partial_r [\alpha r^2 X^{-2} F_r]$ and
$\partial_r [\alpha r^2 X^{-2} P_{rr}]$, as well as the energy flux
terms, $\partial_\epsilon [\epsilon (R_t + O_t)]$ and
$\partial_\epsilon [\epsilon (R_r + O_r)]$ are computed at the
beginning of the neutrino radiation update using the $(n)$ time step
values of $E$, $F_r$, and $P_{rr}$. Treating these flux terms as
explicit, we then solve \esref{eq:e} and \ref{eq:fr} implicitly for
$E^{(n+1)}$ and $F_r^{(n+1)}$ via,
\begin{eqnarray}
  \nonumber 
  \frac{E^{(n+1)} - E^{(n)}}{\Delta t}&=&- \frac{1}{r^2}\partial_r [\frac{\alpha r^2}{X^2}F_r^{(n)}] -
  \partial_\epsilon [\epsilon (R_t^{(n)} + O_t^{(n)})]\\
&& + G_t^{(n+1)} +C_t^{(n+1)}\,, \\
\nonumber 
\frac{F_r^{(n+1)} - F_r^{(n)}}{\Delta t}&=&- \frac{1}{r^2}\partial_r [\frac{\alpha r^2}{X^2}P_{rr}^{(n)}] -
  \partial_\epsilon [\epsilon (R_r^{(n)} + O_r^{(n)})]\\
&& + G_r^{(n+1)} +C_r^{(n+1)}\,.
\end{eqnarray}
The explicit treatment of the flux calculations deserves special
comment and will be discussed in detail in \sref{sec:encoupling} and
\sref{sec:exp_flux}. The source terms will be presented in
\sref{sec:sources} along with the procedure for updating the
conservative hydrodynamic variables at the end of the radiation step.

\subsection{Solving for Higher Moments}
\label{sec:closure}

For any moment expansion, the system of moment equations must be
closed by assuming a closure relation for higher moments. In the
evolution equation for the first moment, \eref{eq:fr}, $P_{rr}$
represents the second moment of the lab-frame neutrino distribution
function. We choose to close the system of equations here and must
specify $P_{rr}$ via a closure relation. Other second moments,
$P^\theta_{\ \theta}$ and $P^\phi_{\ \phi}$, as well as some third moments
$\mathcal{W}^{rrr}$, $\mathcal{W}^{r\theta}_{\ \ \theta}$, and
$\mathcal{W}^{r\phi}_{\ \ \phi}$ are also present in the geometric source
term $G_{r}$ (\sref{sec:geosourceterms}) and energy flux terms $R_\alpha$
and $O_\alpha$ (\sref{sec:encoupling}). We
present the methods for solving for higher moments of the neutrino
distribution function here. We closely follow the work of
\cite{cardall:12rad}.

The determination of these higher moments is most easily done in the
fluid frame rather than in the lab frame. This allows one to ignore
contributions to the neutrino momentum from background motions of the
fluid when performing the closure itself. To determine the fluid frame
moments we use the neutrino stress energy tensor, $T^{\mu \nu}$. The
fluid frame moments ($\mathcal{J}$, $\mathcal{H}^\mu$,
$\mathcal{K}^{\mu \nu}$) are the components of $T^{\mu \nu}$ when it
is expressed in a frame tied to an observer moving with a
four-velocity of $u^\mu$ (i.e.\ with the four-velocity of the fluid; in
\code{GR1D}, $u^\mu = [W/\alpha,Wv^r,0,0]$).  Then,
\begin{eqnarray}
\nonumber
\mathcal{J} & = & u_\mu u_\nu T^{\mu \nu}\,,\\
\nonumber
\mathcal{H}^\mu & = & -u_\nu h^\nu_\rho T^{\mu \rho}\,, \\
\mathcal{K}^{\mu \nu} & = &h^\mu_\rho h^\nu_\sigma T^{\rho \sigma}\,,\label{eq:fluidmoments}
\end{eqnarray}
where $h_{\alpha \beta} = g_{\alpha \beta} + u_\alpha u_\beta$ is the
projection operator. A similar projection of $T^{\mu \nu}$ into the
frame of an observer at rest with respect to the lab frame coordinates
yields the lab-frame moments. Such an observer would have a
four-velocity of $n^\alpha$ (For completeness, in \code{GR1D}, due to
the gauge choice of $\beta^i = 0$, the components of $n^\alpha$ in the
lab frame are: $n^t = 1/\alpha$, $n^i = 0$). The lab frame moment then
follow as,
\begin{eqnarray}
\nonumber
E & = & n_\mu n_\nu T^{\mu \nu}\,,\\
\nonumber
F^\mu & = & -n_\nu \gamma^\nu_\rho T^{\mu \rho}\,, \\
P^{\mu \nu} & = &\gamma^\mu_\rho \gamma^\nu_\sigma T^{\rho \sigma}\,,\label{eq:labmoments}
\end{eqnarray}
where $\gamma_{\alpha \beta}$ is the spatial part of the full metric
$g_{\alpha \beta}$.  The neutrino stress energy tensor can be
constructed from the zeroth, first, and second moments in any
frame. In the lab frame:
\begin{equation}
  T^{\mu \nu} = E n^\mu n^\nu + F^\mu n^\nu + F^\nu n^\mu + P^{\mu 
  \nu}\,, \label{eq:labSET}
\end{equation}
and in the fluid frame,
\begin{equation}
T^{\mu \nu} = \mathcal{J} u^\mu u^\nu + \mathcal{H}^\mu u^\nu + \mathcal{H}^\nu u^\mu + \mathcal{K}^{\mu 
  \nu}\,. \label{eq:fluidSET}
\end{equation}
The general procedure for determining the higher moments is:
\emph{1)} Use the lab-frame energy ($E$) and momentum density ($F_r$)
along with a guess for the lab-frame second moment ($P_{rr}$; usually
the previous value) to construct the neutrino stress energy tensor for
a particular energy group via \eref{eq:labSET}. \emph{2)} Determine
the fluid frame moments via \eref{eq:fluidmoments}. \emph{3)} Use the
analytic closure below to determine the fluid-frame second moment from
the zeroth and first moments $\mathcal{K}^{\mu
  \nu}(\mathcal{}J,\mathcal{H}^\mu)$ \emph{4)}
Reconstruct the stress energy tensor, now with the fluid frame moments
via \eref{eq:fluidSET}, and \emph{5)} project out the lab-frame second
moments via \eref{eq:labmoments}. Since the lab frame second moment,
$P_{rr}$, entered into the original stress energy tensor, we must
iterate this process until we reach convergence on the lab-frame
second moment.

The analytic closure we apply to determine $\mathcal{K}^{\mu \nu}$ comes from
\cite{cardall:12rad} Eq.~106,
\begin{equation}
  \mathcal{K}^{\mu \nu} = \frac{\mathcal{J}}{3}h^{\mu \nu} + a(\mathcal{J},\mathcal{H}^2)\left(\mathcal{H}^\mu \mathcal{H}^\nu - \frac{\mathcal{H}^2}{3}h^{\mu \nu}\right)\,,
\end{equation}
where we take $a(\mathcal{J},\mathcal{H}^2) =
\mathcal{J}/\mathcal{H}^2 \times (3\chi-1)/2$ with $\mathcal{H}^2 =
\mathcal{H}^\mu \mathcal{H}_\mu$. This choice of
$a(\mathcal{J},\mathcal{H}^2)$ gives the more commonly found form of
of the second moment,
\begin{equation}
\mathcal{K}^{\mu \nu} = \frac{3(1-\chi)}{2} \mathcal{K}_\mathrm{thick}^{\mu \nu} +
\frac{3\chi - 1}{2} \mathcal{K}_\mathrm{thin}^{\mu \nu}\,,
\end{equation}
where 
\begin{equation}
\mathcal{K}_\mathrm{thick}^{\mu \nu} = \frac{\mathcal{J}}{3}h^{\mu\nu}\,,
\end{equation}
is the analytic second moment derived from the diffusion limit (where
the radiation field is isotropic) and
\begin{equation}
\mathcal{K}_\mathrm{thin}^{\mu \nu} = \mathcal{J}\frac{\mathcal{H}^\mu \mathcal{H}^\nu}{\mathcal{H}^2}\,.
\end{equation}
is the free streaming limit. $\chi$ in these equations plays the role
of an interpolation factor.  In the optically thick limit, it
asymptotes to 1/3 (giving $\mathcal{K}^{\mu
  \nu}=\mathcal{K}_\mathrm{thick}^{\mu \nu}$) and in the free
streaming limit it approaches 1 (giving $\mathcal{K}^{\mu
  \nu}=\mathcal{K}_\mathrm{thin}^{\mu \nu}$). By default, \code{GR1D}
uses the common choice of the Minerbo closure
\citep{minerbo:78,pons:00},
\begin{equation}
\chi = \frac{1}{3} + \frac{2}{15}\left(3f^2-f^3+3f^4\right)\,; f = \left( \frac{\mathcal{H}^2}{\mathcal{J}^2} \right)^{1/2}\,,\label{eq:closure}
 \end{equation}
 where $f$ is the flux factor and is a proxy for the forward
 peakedness of the distribution function. When $f$ is zero the
 radiation field is isotropic. When $f$ is one, the radiation field is
 completely forward peaked in the direction of the neutrino momentum.

 For the analytic fluid frame third moments
 ($\mathcal{L}^{\mu \nu \rho}$) we use, Eq.~108 from
 \cite{cardall:12rad} taking
 $b(\mathcal{J},\mathcal{H}) = 1/\mathcal{H}^2 \times (3\chi-1)/2$,
\begin{equation}
\mathcal{L}^{\mu \nu \rho} = \frac{3(1-\chi)}{2} \mathcal{L}_\mathrm{thick}^{\mu \nu \rho} +
\frac{3\chi - 1}{2} \mathcal{L}_\mathrm{thin}^{\mu \nu \rho}\,,
\end{equation}
where
\begin{equation}
\mathcal{L}_\mathrm{thick}^{\mu \nu \rho} = \frac{1}{5}\left(\mathcal{H}^\mu h^{\nu\rho} +
  \mathcal{H}^\rho h^{\mu\nu} + \mathcal{H}^\nu h^{\mu\rho}\right) \,,
\end{equation}
and
\begin{equation}
\mathcal{L}_\mathrm{thin}^{\mu \nu \rho} = \frac{\mathcal{H}^\mu \mathcal{H}^\nu \mathcal{H}^\rho}{\mathcal{H}^2}\,.
\end{equation}
From this, we compute the lab frame third moments, denoted here as
$\mathcal{W}^{\mu \nu \rho}$, via Eq.~B18 of \cite{cardall:12rad}. The
moments relevant for \code{GR1D} are
\begin{equation}
\mathcal{W}^{rrr} = \mathcal{L}^{rrr} +  3\left(\frac{Wv}{X}\right)\mathcal{K}^{rr} +
3\left(\frac{Wv}{X}\right)^2\mathcal{H}^r + \left(\frac{Wv}{X}\right)^3 \mathcal{J}\,,
\end{equation}
and  
\begin{equation}
  \mathcal{W}^{r\phi}_{\ \ \phi} = \mathcal{L}^{r\phi}_{\ \ \phi} + \left(\frac{Wv}{X}\right)\mathcal{K}^\phi_{\ \phi}\,.
\end{equation}

\subsection{Methods for Coupling Energy Groups}
\label{sec:encoupling}

The components of the energy flux terms, $R_\alpha$ and $O_\alpha$
from \esref{eq:e} and \ref{eq:fr}, are given in \cite{cardall:12rad}
and reproduced here using the metric and notation of \code{GR1D}.
\begin{eqnarray}
  R_t &=& \alpha W\left[ (\frac{\mathcal{Z}v}{X} -
  \mathcal{Y}^r)\partial_r \phi -
  \frac{\mathcal{X}^{ii}v}{2X}\partial_r g_{ii} +
  \mathcal{X}^{rr}K_{rr}\right]\,,\label{eq:rt}\\
  O_t &=& \alpha \left [\frac{\mathcal{Z}}{\alpha}\partial_t
  W + \mathcal{Y}^r\partial_r W -
  \frac{\mathcal{Y}_r}{\alpha}\partial_t\frac{Wv}{X} - \mathcal{X}_r^r\partial_r\frac{Wv}{X}\right]\,,\label{eq:ot}
\end{eqnarray}
and
\begin{eqnarray}
R_r &=& \alpha W \left[
(\frac{\mathcal{Y}_rv}{X}-\mathcal{X}_r^r)\partial_r \phi -
\frac{\mathcal{W}_r^{\ ii}v}{2X}\partial_rg_{ii} +
\mathcal{W}_r^{rr}K_{rr}\right]\,,\label{eq:rr}\\
O_r &=& \alpha
\left[\frac{\mathcal{Y}_r}{\alpha}\partial_tW +
  \mathcal{X}_r^r\partial_rW
  -\frac{\mathcal{X}_{rr}}{\alpha}\partial_t\frac{Wv}{X} - \mathcal{W}_{rr}^r\partial_r\frac{Wv}{X}\right]\,,\label{eq:or}
\end{eqnarray}
where
\begin{eqnarray}
W \mathcal{Z} &=& E + \frac{v F_r}{X} + \frac{v^2 P_{rr}}{X^2} +
\frac{W v^3}{X^3} \mathcal{W}_{rrr}\,,\\
W \mathcal{Y}^r &=& \frac{F_r}{X^2} + \frac{v P_{rr}}{X^3} +
\frac{W v^2}{X^4} \mathcal{W}_{rrr}\,,\\ 
W \mathcal{X}^{rr} &=& \frac{P_{rr}}{X^4} +
\frac{W v}{X^5} \mathcal{W}_{rrr}\,,\\ 
W \mathcal{X}^{\theta \theta} &=& \frac{P_\theta^{\ \theta}}{r^2} +
\frac{W v}{X r^2} \mathcal{W}_{r\theta}^{\ \ \theta}\,,\\ 
W \mathcal{X}^{\phi \phi} &=& \frac{P_\phi^{\ \phi}}{r^2} +
\frac{W v}{X r^2} \mathcal{W}_{r\phi}^{\ \ \phi}\,.
\end{eqnarray}
$K_{rr} = -X\dot{X}/\alpha$ (not to be confused with
$\mathcal{K}^{\mu \nu}$) in \esref{eq:rt} and \ref{eq:rr} is the
extrinsic curvature. Required radial and time derivatives of the
metric quantities are analytic in the metric of \code{GR1D} and
available in \cite{oconnor:10}. For the radial and time derivatives of
the velocity we use finite differencing,
$\dot{v} = (v^{(n+1)}-v^{(n)})/(t^{(n+1)} - t^{(n)})$ and
$dv_i/dr = (v_{i+1} - v_{i-1})/(x_{i+1} - x_{i-1})$.

For determining the inter-group fluxes we follow the number-conserving
scheme of \cite{mueller:10}. This scheme computes the momentum space
fluxes via the equations presented here and reconstructs the flux at
the energy group interface by assigning weights to the left and right
states. The weights are determined in such a way as to conserve
neutrino number. \code{GR1D} can treat these energy group couplings
implicitly or explicitly. Due to the small time step enforced via the
spatial flux treatment (see the following section), we find an
explicit treatment of these terms is sufficient for typical situations
and avoids a large matrix inversion.

\subsection{Explicit Update for the Spatial Flux}
\label{sec:exp_flux}

The transport of neutrinos from one spatial zone to another is handled
via the spatial flux term in \esref{eq:e} and \ref{eq:fr}. We solve
these terms with the standard hyperbolic methods used for conservative
hydrodynamics and apply asymptotic solutions for the optically thick
regimes where the hyperbolic methods fail.\footnote{We are indebted to
  Luke Roberts for many discussions on this topic and for the ultimate
  $\mathcal{O}(v/c)$ solution in \code{GR1D}.}

We base the spatial flux term calculation on the methods used in the
HLLE \citep{HLLE:88} Riemann solver. First, the lab-frame moments are
reconstructed to the left and right sides of the cell interface. In
\code{GR1D} we adopt the same reconstructor we use for the
hydrodynamics, either TVD (for the collapse phase) or piecewise
parabolic (once the density has reached $10^{12}$\,g\,cm$^{-3}$). For
non core collapse test cases in this paper, we use TVD.  In practice,
we reconstruct the zeroth moment ($E$) and the ratio of the first
moment to the zeroth moment ($F_{r}/E$) to insure that $F_{r}/E$ at
the interface never exceeds the value in the zone center and therefore
remains casual.  The closure is re-solved to obtain the interface
values of the second moment in the lab-frame. Following
\cite{shibata:11}, we estimate the characteristic speeds needed in the
Riemann solution via an interpolation between the optically thick and
free streaming regimes,
\begin{equation}
\lambda^\mathrm{max/min} = \frac{3(\chi-1)}{2}\lambda^\mathrm{max/min}
_\mathrm{thick} + \frac{3\chi-1}{2}\lambda^\mathrm{max/min}
_\mathrm{thin}\,,
\end{equation}
where $\chi$ is computed as part of the closure (see
\sref{sec:closure}),
\begin{equation}
\lambda^\mathrm{max/min}_\mathrm{thick} = \mathrm{max/min}(\alpha 
X\frac{2W^2v\pm\sqrt{3}}{2W^2+1},\alpha X v)\,,
\end{equation}
and
\begin{equation}
\lambda^\mathrm{max/min}_\mathrm{thin} = \mathrm{max/min}(\pm \alpha X)\,.
\end{equation}
The inter-cell fluxes from the Riemann solution are then given as
\begin{equation}
  F^{i+1/2,\mathrm{HLLE}}_r = \frac{\lambda^\mathrm{max}F^{i,R}_r -
  \lambda^\mathrm{min}F^{i+1,L}_r +
  \lambda^\mathrm{max}\lambda^\mathrm{min}(E^{i+1,L} -
  E^{i,R})}{\lambda^\mathrm{max} - \lambda^\mathrm{min}}\,,
\end{equation}
and
\begin{equation}
P^{i+1/2,\mathrm{HLLE}}_{rr} = \frac{\lambda^\mathrm{max}P^{i,R}_{rr} -
  \lambda^\mathrm{min}P^{i+1,L}_{rr} +
  \lambda^\mathrm{max}\lambda^\mathrm{min}(F_r^{i+1,L} -
  F_r^{i,R})}{\lambda^\mathrm{max} - \lambda^\mathrm{min}}\,,
\end{equation}
where $A^{i,R/L}$ are the reconstructed moments to the right/left
interface in zone $i$. In the optically thick regime, or what we refer
to as the high Peclet number regime (Pe $\sim (\Delta x_i \times
\kappa_i)$; or the optical depth of the zone $i$), the diffusive term
in the Riemann solution becomes dominated by numerical noise and is no
longer accurate \citep{audit:02}. In these regions we replace the
interface fluxes by their asymptotic values. In
\code{GR1D}, the $\mathcal{O}(v/c)$ approximation for this asymptotic
flux is \citep{roberts:14priv}

\begin{equation}
F^{i+1/2,\mathrm{asym}}_r = \frac{4W^2vX}{3}\mathcal{J} -
\frac{W}{3\bar{\kappa}X^2}\frac{\partial \mathcal{J}}{\partial r}\,,
\end{equation}
where the first term is the flux due to advection with the fluid, and
the second term is the flux due to diffusion. For the advection we use
an explicit upwind scheme. We estimate $\partial J/\partial r$ via a
simple finite difference of the fluid frame energy densities. For the
momentum flux in the high Peclet number regime, we take a simple
average of the neighboring zone's second moment for the asymptotic
flux,
\begin{equation}
P^{i+1/2,\mathrm{asym}}_{rr} = (P^{i}_{rr} + P^{i+1}_{rr})/2\,.
\end{equation}

We interpolate between the two regimes using the Peclet
number. Following \cite{jin:96,audit:02}, 
\begin{equation}
a = \tanh(1/\overline{\mathrm{Pe}})\,,\label{eq:peclet}
\end{equation}
where $\overline{\mathrm{Pe}} =
\sqrt{(\kappa^i_s+\kappa^i_a)(\kappa^{i+1}_s+\kappa^{i+1}_a)}(x_{i+1}-x_i)W^3(1+v)X^2$
and $\kappa_s$ and $\kappa_a$ are the scattering and absorption
opacities, respectively. The extra $W^3(1+v)X^2$ terms arise from the
coefficients of the neutrino momentum sink term in $C_r$ (see
\sref{sec:sources} and \citealt{audit:02}). $a=1$ in regions of small
Peclet number, $a \propto 1/\mathrm{Pe}$ in regions with large Peclet
number, and the $\tanh$ function gives a smooth interpolation in
between. The ultimate value for the fluxes on the cell interfaces is,
\begin{equation} F^{i+1/2}_r = a\times F^{i+1/2,\mathrm{HLLE}}_r +
  (1-a)\times F^{i+1/2,\mathrm{asym}}_r \,,\label{eq:energyflux}
\end{equation}
and
\begin{equation} P^{i+1/2}_r = a\times
  P^{i+1/2,\mathrm{HLLE}}_{rr} + (1-a)\times
  P^{i+1/2,\mathrm{asym}}_{rr} \,. \label{eq:momentumflux}
\end{equation}

The flux update terms in the moment evolution equations are taken to be
\begin{equation}
\partial_r [\frac{\alpha r^2}{X^2}F_r^{(n)}] = \frac{1}{\Delta r^i}\left\{\left[\frac{\alpha 
      r^2}{X^2}\right]^{i+1/2}F^{i+1/2}_r - \left[\frac{\alpha r^2}{X^2}\right]^{i-1/2}F^{i-1/2}_r\right\}\,,
\end{equation}
and
\begin{equation}
\partial_r [\frac{\alpha r^2}{X^2}P_{rr}^{(n)}] = \frac{1}{\Delta r^i}\left\{\left[\frac{\alpha 
      r^2}{X^2}\right]^{i+1/2}P^{i+1/2}_{rr} - \left[\frac{\alpha r^2}{X^2}\right]^{i-1/2}P^{i-1/2}_{rr}\right\}\,.
\end{equation}

The CFL condition restricts the maximum time step that can be taken to
the light crossing time of the smallest spatial zone. Additionally, we
reduce the time step via a Courant factor of 0.5. In \code{GR1D}, this
sets the time step for both the hydrodynamic step and the neutrino
radiation step. While the hydrodynamic step uses a second-order
Runge-Kutta for the time evolution, the neutrino radiation step uses a
simpler first order scheme for both the implicit and explicit parts.
While first-order explicit methods are normally not used because they
are very inaccurate, two aspects alleviate this issue in \code{GR1D}.
\code{GR1D} generally uses a logarithmically spaced grid. The
innermost zones are $\mathcal{O}(200m)$ and these set the time step.
However, in these zones the radiation is generally optically thick and
therefore the fluxes do not have characteristic speeds close to the
speed of light. For the free streaming regions farther out, the speeds
are close to the speed of light, however, due to the logarithmic
spacing, these zones are much larger than the smallest zone. The only
place where there is nearly free streaming radiation closer to the
innermost zones is near bounce when the extent of the supersonic flow
reaches down to $\sim10\,$km. For this reason, and because the epoch
of core bounce is very dynamic, we decrease the Courant factor to 0.25
when the central density first reaches $10^{12}$\,g\,cm$^{-3}$. We
increase the Courant factor back to 0.5 at 20\,ms after core bounce.
At times other than near core bounce, we do not see differences when
decreasing the Courant factor below 0.5. This give us confidence that
this treatment is sufficient for accurate evolutions.

\subsection{Source terms}
\label{sec:sources}

The neutrino radiation fields are sourced and sinked by weak
interaction processes occurring in the matter, between the neutrinos
and the matter, or between the neutrinos themselves. There are also
geometric source terms that arise due to our particular set of
evolution equations. \code{GR1D} can currently handle most types of
standard neutrino-matter interactions. In this section, we will
discuss these types of interactions and how they are included in both
the neutrino and hydrodynamic evolution equations. In the following
section we introduce \code{NuLib}, an open-source neutrino interaction
library and describe the specific neutrino-matter interactions it
includes.

\subsubsection{Geometric Source Terms}
\label{sec:geosourceterms}

The geometric source terms, $G_\alpha$ in \esref{eq:e} and \ref{eq:fr}
are given in both \cite{cardall:12rad} and \cite{shibata:11}, we
repeat them here in the notation and metric of \code{GR1D}. We note
that the differences in the definition of the evolved variables
between \code{GR1D} and these other works affects the definitions of
$G_t$ and $G_r$ (i.e. \code{GR1D} evolves
$E$ and $F_r$ as opposed to the densitized versions,
$\sqrt{\gamma}E$ and $\sqrt{\gamma}F_r$ where $\gamma$ is the
determinant of the metric). The geometric source term for the zeroth neutrino
moment is
\begin{equation}
G_t = \alpha 4\pi r\rho h W^2\left[ E v X (1+p_{rr}/X^2) - F_r ( 1+v^2)\right]\,,
\end{equation}
where $p_{rr} = P_{rr}/E$ is the Eddington factor and $h$ is the
enthalpy defined in \sref{sec:GRhydro}. While the geometric source
term for the momentum density is
\begin{eqnarray}
\nonumber
G_r & = &  -\alpha 4 \pi r \rho h W^2 F_r v X - \alpha E X^2 \left[ \frac{m}{r^2} + 4 \pi r (P+\rho h W^2
  v^2)\right]  \\
&& + \alpha E \frac{p^\phi_\phi+p^\theta_\theta}{r}\,,
\end{eqnarray}
where $p^i_{\ i} = P^i_{\ i}/E$ and we note that $P$ is the matter
pressure and $m$ is the enclosed mass given by \eref{eq:mass}.

\subsubsection{Neutrino Source Terms}
\label{sec:nusourceterms}

For the neutrino-matter interaction source terms we follow the source
term formalism of \cite{shibata:11}. The neutrino source terms in
\esref{eq:e} and \ref{eq:fr} are,
\begin{equation}
C_t = -\alpha n_\alpha \left[ S^\alpha_\mathrm{e/a} +
  S^\alpha_\mathrm{iso} + S^\alpha_\mathrm{scatter} + S^\alpha_\mathrm{thermal}\right]\,,\label{eq:neutrinosourcest}
\end{equation}
and
\begin{equation}
C_r = \alpha \gamma_{r\alpha} \left[ S^\alpha_\mathrm{e/a} +
  S^\alpha_\mathrm{iso} + S^\alpha_\mathrm{scatter}  + S^\alpha_\mathrm{thermal}\right]\,,\label{eq:neutrinosourcesr}
\end{equation}
where $S^\alpha_\mathrm{e/a}$ is the source term for emission and
absorption of neutrinos from and into the matter, respectively,
$S^\alpha_\mathrm{iso}$ is the source term for elastic scattering of
neutrinos off the matter, $S^\alpha_\mathrm{scatter}$ encompasses
the source term for inelastic scattering of neutrinos off matter,
and $S^\alpha_\mathrm{thermal}$ describes thermal production of
neutrino-antineutrino pairs and their annihilation. These terms are
expressed most easily in the fluid rest frame and will be functions of
$\mathcal{J}$, $\mathcal{H^\alpha}$, $\mathcal{K^{\alpha \beta}}$, the
fluid four-velocity $u^\alpha$, and the neutrino-matter interaction
coefficients. For solving the implicit step, it is necessary to know
the neutrino source terms in terms of the lab frame moments $E$,
$F_r$, and the closure relations of \sref{sec:closure}. Using
\esref{eq:fluidmoments} and \ref{eq:labSET}, it is easily possible to show
that in \code{GR1D},
\begin{equation}
\mathcal{J} = W^2\left[E - 2F_r v /X + P_{rr}v^2 / X^2\right]\,,\label{eq:fluidJ}
\end{equation}
and 
\begin{eqnarray}
\mathcal{H}^t &=& \frac{W^3}{\alpha}\left[-(E - F_rv/X)v^2 + vF_r/X - v^2P_{rr}/X^2\right]\,,\label{eq:ht}\\
\mathcal{H}^r &=& W^3\left[-(E - F_rv/X)v/X + F_r/X^2 - vP_{rr}/X^3\right]\,.\label{eq:hr}
\end{eqnarray}
When we perform the implicit solve for $E^{(n+1)}$ and $F_r^{(n+1)}$
we assume $P^{(n+1)}_{rr} = p_{rr}^{(n)} E^{(n+1)}$. Since we first
solve the hydrodynamic step before the radiation step, we have an
estimate of $\rho^{(n+1)}$, $T^{(n+1)}$, and $Y_e^{(n+1)}$. We use
these to determine the neutrino interaction terms. We do not include
$T$ and $Y_e$ in the implicit solution (and therefore do not
recalculate these interaction coefficients when finding the solution),
rather we update them once after we solve for $E^{(n+1)}$ and
$F_r^{(n+1)}$. These approximations are justified since, unlike most
implicit schemes, we are limited to very small time steps from the
explicit spatial flux calculation (see \sref{sec:exp_flux}).

From \cite{shibata:11},
\begin{equation}
S^\alpha_\mathrm{e/a} = \left[ \eta - \kappa_a \mathcal{J}\right]
u^\alpha - \kappa_a \mathcal{H}^\alpha\,,
\end{equation}
where $\eta$ is the emissivity and $\kappa_a$ is the absorption
opacity for neutrinos of a given species and energy. Weak interactions
that produce neutrinos (such as electron capture on protons) and
destroy neutrinos (such as the inverse interaction of electron
neutrino capture on neutrons) are used to compute $\eta$ and
$\kappa_a$. We describe this in more detail in \sref{sec:nulib} on
\code{NuLib}.
\begin{equation}
S^\alpha_\mathrm{iso} = -\kappa_s \mathcal{H}^\alpha\,,\label{eq:siso}
\end{equation}
where $\kappa_s$ is the scattering opacity for neutrinos of a given
species and energy. Only isoenergetic (elastic) interactions where
the neutrino survives the interaction with the same energy as it
started will contribute to this scattering opacity.

These first two neutrino source terms were mono-energetic. They do not
depend on the neutrino distribution function of other neutrino species
or energies. $S^\alpha_\mathrm{scatter}$ is more complicated since it
depends on the neutrino distribution function at all other neutrino
energies. The inelastic neutrino-matter scattering opacity depends on
the occupation level of the final state neutrino. If that energy level
is completely filled with other neutrinos than the scattering is
blocked. Therefore, at a given energy, we must consider the scattering
to every other neutrino energy individually and take into account the
neutrino phase space occupancy of that energy group. This process
couples all of the energy bins for a given species and can
dramatically increase the required computational resources if many
energy groups are considered.  To avoid a mere restatement of complex
formulas, we refer the reader to \cite{shibata:11} for specific
definitions of $S^\alpha_\mathrm{scatter}$ (their Eq.~4.14) but note
the slight notation differences between our work and their's \emph{a)}
their $L^{\alpha \beta}$ is our $\mathcal{K}^{\alpha \beta}$),
\emph{b)} their fluid frame moments are also integrated over solid
angle and therefore are $4\pi$ larger than ours. We discuss the
scattering kernels ($R_{0/1}^\mathrm{in/out}$) calculation in
\code{NuLib} in the following section.

Since the full implicit calculation of inelastic scattering is
computationally demanding, any potential simplification and/or
approximation is greatly desired. In the context of CCSN simulations
in \code{GR1D}, we have found that we can forgo the full implicit
inelastic neutrino-electron scattering calculation and instead include
this interaction as an explicit term in \esref{eq:e} and \ref{eq:fr}.
This works in part because of the small time step afforded to us from
the explicit flux calculation. However, in order to achieve a stable
evolution when the time scale of inelastic scattering is shorter than
our time step, we must decrease the magnitude of the scattering
kernels that enter into $S^\alpha_\mathrm{scatter}$. Following
inspiration from \cite{thompson:03}, we have empirically found that
suppressing the scattering kernels at densities above
$\rho = 5\times10^{12}$\,g\,cm$^{-3}$ via,
\begin{equation}
R_{0/1}^{in/out,*} = R_{0/1}^{in/out}
/\mathrm{max}[1,(\rho/(5\times 10^{12}\,\mathrm{g}\,\mathrm{cm}^{-3})^{3/2})]\,,
\end{equation}
is sufficient. This effectively slows down the scattering of the
neutrinos in these high density regimes and will likely be invalid when
the neutrinosphere reaches these densities. In \sref{sec:approxs}, we
show that there are no serious side effects of this approximation for
a typical early-phase CCSN in \code{GR1D}.

The last main group of neutrino interactions are neutral current
pair-process interactions or thermal interactions. They enter into the
term $S_\mathrm{thermal}^\alpha$ in \esref{eq:neutrinosourcest} and
\ref{eq:neutrinosourcesr}. These are processes that produce
neutrino-antineutrino pairs through, for example, electron-positron
annihilation. A fully consistent treatment of these processes requires
coupling neutrino species as well as energy groups.  We leave this for
future work and treat these processes in \code{GR1D} via an
approximation\footnote{However, we carry out several simulations in
  \sref{sec:approxs} where we treat $\nu_x \bar{\nu}_x$
  production/annihilation via electron-positron
  annihilation/production fully (via \cite{shibata:11}'s methods) to
  test this approximation. As we do not yet do this for electron type
  neutrinos, or for other processes like nucleon-nucleon
  Bremsstrahlung, we leave a formal discussion of it to future work.}. We
approximate the neutrino-antineutrino annihilation rate via an
interpolation between the limiting regimes. We assign an absorption
opacity to this interaction via $\kappa_a^{pp} = \eta^{pp} / B_\nu$
where $\eta^{pp}$ is the isotropic emissivity of neutrinos due to some
thermal process (i.e. electron-positron annihilation) assuming no
final state neutrino blocking and $B_\nu$ is the neutrino black-body
function for that density, temperature and electron fraction. This
ensures both the correct emission rate when there are no neutrinos
present to block neutrino-antineutrino production and it ensures that
in equilibrium (when $\mathcal{J} = B_\nu$) there is no net emission
since $\eta^{pp} - \kappa^{pp}_a \mathcal{J}$ will be zero by
construction. By default, we use this approximation for producing and
`annihilating' $\nu_x$ neutrinos in \code{GR1D}. Furthermore, we
ignore thermal processes for electron type neutrino and antineutrinos
because the production of these neutrinos is dominated by charged
current processes over pair-production processes. For CCSNe, we find
that our heavy-lepton production/annihilation approximation is highly
successful giving comparable neutrino signals to an implicit
solution. (see \sref{sec:s15compare} and \sref{sec:approxs} for
details, but we find luminosities within $\sim$7\% and root mean
squared energies within $\sim$2\%).

\subsubsection{Neutrino-Matter coupling}
\label{sec:numattercoupling}

The neutrino-matter source terms are also source terms for the matter
evolution equations. They can influence the energy content, the lepton
content, and the momentum of the matter. We update the conservative
hydrodynamic variables (from \eref{eq:chv}) at the end of the neutrino
radiation step with the corresponding change to the neutrino
variables\footnote{This is different from most fully implicit methods
where the matter coupling is also included in the implicit step. Since
our time step is forced to be small by the light crossing time of the
smallest zone, we find that coupling the matter into the implicit
source term step is not necessary.}. This ensures any energy, leptons,
or momentum sourced into the neutrino fields is subtracted from the
matter.
\begin{eqnarray}
\Delta [\tau] &=& -\Delta t \times 4 \pi \alpha^2 \sum_{\epsilon,\nu_i} S^t\,,\label{eq:dtau}\\
\Delta [S_\mathrm{hydro}^r] &=& -\Delta t \times 4 \pi \alpha X \sum_{\epsilon,\nu_i} S^r\,,\label{eq:dsr}\\
\Delta [D Y_e] &=& -\Delta t \times 4 \pi \alpha X m_N \sum_{\epsilon,\nu_i}
s_{\nu_i} W\left[S^t \alpha - S^r X v\right]/\epsilon\,,\label{eq:dye}
\end{eqnarray}
where $\Delta t$ is the numerical time step, $S^t$ and $S^r$ are the
source terms in \esref{eq:neutrinosourcest} and
\ref{eq:neutrinosourcesr} evaluated in the lab frame and $s_{\nu_i}$
denotes lepton number: 1 for $\nu_e$, -1 for $\bar{\nu}_e$, and 0 for
$\nu_x$. For $\Delta [\tau]$ and $\Delta [S_\mathrm{hydro}^r]$, the
preceding factors of $\alpha^2$ and $\alpha X$ arise from
\esref{eq:neutrinosourcest} and \ref{eq:neutrinosourcesr},
respectively. We note that there is a subtle factor of $X$ between the
definitions of the conserved hydro quantity $S^r_\mathrm{hydro}$ and
the neutrino momentum $F_r$ such that the sum of $\Delta
[S^r_\mathrm{hydro}X + F_r] = 0$ (see Appendix A of
\citealt{oconnor:10} for details of the precise definition of
$S^r_\mathrm{hydro}$). Since the neutrino energies are defined in the
fluid rest frame, the lepton number change must be computed from the
energy source term in that frame (i.e. rather than taking $s_{\nu_i}
S^t/\epsilon$ as the lepton source term). Therefore, for the
definition of $\Delta [D Y_e]$ in \eref{eq:dye}, the calculation could
also be carried out with the more intuitive $[\eta - \kappa_a
\mathcal{J}]/\epsilon$ instead of the more cumbersome $W[S^t \alpha -
S^r X v]/\epsilon$. These expressions are equivalent as can be shown
via \eref{eq:ht} - \eref{eq:siso}. Finally, we note that
$S^\alpha_\mathrm{scatter}$ and $S^\alpha_\mathrm{thermal}$ will
contribute to both \eref{eq:dtau} and \eref{eq:dsr}, but not to
\eref{eq:dye}.

The other neutrino-matter coupling is through the gravitational field
and we are now in a position to derive the neutrino contributions to
the metric equations alluded to in \esref{eq:mass} and \eref{eq:phi}.
The neutrino stress energy tensor is simply the energy integrated form
of \eref{eq:labSET} or \eref{eq:fluidSET},
\begin{equation}
T^{\mu \nu}_\mathrm{neutrino} = 4\pi\sum_{\mathrm{species}\,i}\,\,
\sum_{\mathrm{energy}\,\epsilon} T^{\mu \nu}_{i,\epsilon}
\Delta \epsilon\,.
\end{equation}
While both give the correct contribution, we choose to use the
lab-frame representation of the neutrino stress energy tensor,
\eref{eq:labSET}. This gives \citep{gourgoulhon:91},
\begin{eqnarray}
\tau_m^\nu &=& -{T^t_t}_\mathrm{neutrino} =4\pi \sum_{\mathrm{species}\,i}\,\,
\sum_{\mathrm{energy}\,\epsilon} E_{i,\epsilon} \Delta \epsilon\,,\\
\tau_\phi^\nu &=& -{T^r_r}_\mathrm{neutrino} =  4\pi \sum_{\mathrm{species}\,i}\,\,
\sum_{\mathrm{energy}\,\epsilon} P^{rr}_{i,\epsilon} \Delta \epsilon\,.
\end{eqnarray}

\section{Neutrino Microphysics: \code{NuLib}}
\label{sec:nulib}

The neutrino-matter interaction coefficients introduced in
\sref{sec:sources} contain contributions from many different
processes. These coefficients depend directly on the matter density,
$\rho$; temperature, $T$; matter chemical potentials, $\mu_e$,
$\mu_n$, $\mu_p$; and the nuclear isotope distribution and therefore
require an EOS to compute. They also strongly depend on neutrino
energy and species. Similar to what is done for nuclear EOS, for
matter in nuclear statistical equilibrium we can precompute all the
neutrino-matter interaction coefficients and tabulate them for quick,
on the fly \emph{interpolation} rather than the slower on the fly
\emph{computation}. While table interpolation has the benefit of
speeding up the determination of the neutrino interaction rates,
there can be consequences of this if the interpolation does not follow
inherent relationships between quantities, we note a few of these
situations below when discussing various processes.

The collection of routines we use to compute the neutrino-matter
interaction coefficients for \code{GR1D} is called \code{NuLib}.
\code{NuLib} is an open-source library and available as a git
repository at \url{http://www.nulib.org}. We have tagged a release
called `GR1Dv2' to accompany this paper. We summarize the rates
included in the simulations of this paper from \code{NuLib} in
\tref{tab:interactions} and discuss details below. Every rate and
correction in \code{NuLib} is optional and can easily be left out of
the calculation for the total rate. For a full description of the
neutrino interactions and their implementation, please consult
documentation and source code at \url{http://www.nulib.org}. The set
of available interactions is constantly evolving. Contributions to
this community resource are welcome.

\begin{table}
\caption{Neutrino Interactions in \code{NuLib}}
\centering
\begin{tabular}{cc}
\multicolumn{2}{c}{Production} \\
\hline
Charged-Current Interactions & Thermal Processes \\
\hline
$\nu_e + n \to p + e^-$ & $e^- + e^+ \to \nu_x +\bar{\nu}_x$ \\
$\bar{\nu}_e + p \to n + e^+$ & $N + N \to
N + N +\nu_x + \bar{\nu}_x$\\
$\nu_e + (A,Z) \to (A,Z+1) + e^-$ &\\
\\
\multicolumn{2}{c}{Scattering} \\
\hline
Iso-Energetic Scattering & Inelastic 
Scattering \\
\hline
 $\nu + \alpha \to \nu + \alpha$ & $\nu_i +e^- \to \nu_i^\prime +
 e^{-\ \prime}$\\
$\nu_i +p \to \nu_i + p$& \\
$\nu_i +n \to \nu_i + n$ &\\
 $\nu + (A,Z) \to \nu + (A,Z)$ &\\
\end{tabular}
\tablecomments{Neutrino interactions from \code{NuLib} used in this study.
  Production interactions with neutrinos on the left are computed as 
  cross sections, production interactions with neutrinos on the right
  are computed as emissivities. Interactions with $\nu$ are not flavor
  sensitive, while interactions with $\nu_i$ are.  Specific
  interactions that only involve one type of neutrino use the specific
  neutrino flavor. $\nu_x$ is used to denote heavy-lepton neutrinos.}\label{tab:interactions}
\end{table}

{\emph{Absorption Cross Sections:}} \code{NuLib}'s neutrino-nucleon
charged-current interaction cross sections are taken from
\cite{brt:06}. Weak-magnetism and recoil corrections from
\cite{horowitz:02} are applied. \code{NuLib} also includes neutrino
absorption on heavy nuclei via the simple treatment of
\cite{brt:06,bruenn:85}. More complete electron capture rates on heavy
nuclei will be included in a future version of \code{NuLib}. Cross
sections are converted to opacities using the target number densities
from the chosen EOS.  Emissivities are computed via Kirchhoff's Law,
which equates the neutrino emission rate to the absorption rate of an
equilibrium distribution of neutrinos. This does not require the
neutrinos to be in equilibrium and is valid regardless of the
background neutrino field. We note that table interpolation is not
guaranteed to maintain this relationship between the emissivity and
opacity. In practice, in \code{GR1D}, we enforce this relationship by
recomputing the emissivity from the absorption opacity after
interpolation.

{\emph{Thermal Neutrino Pair Production:}} For thermal processes,
\code{NuLib} makes the approximation mentioned in
\sref{sec:nusourceterms}. We compute the emissivity assuming there is
no final state neutrino blocking. For electron-positron annihilation,
this calculation is only a function of the electron chemical potential
and matter temperature since this completely sets the electron and
positron distributions. We compute the emissivity following
\cite{brt:06,bruenn:85} which is based on the earlier work of
\cite{yueh:76}. We compute the zeroth moment of the neutrino
production kernels to get the total energy emission. We assume it is
isotropic. For nucleon-nucleon Bremsstrahlung we also follow
\cite{brt:06} and make the same assumption we make for
electron-positron annihilation. \code{NuLib} also computes the zeroth
and first moment of the neutrino production/annihilation kernels for
electron-positron annihilation. This allows us to test the quality of
our approximation. We present the results of that test in
\sref{sec:approxs}.

{\emph{Elastic Scattering:}} For elastic scattering processes we
include scattering off nucleons, alpha particles, and heavy nuclei
following \cite{bruenn:85,brt:06}. For elastic scattering on nucleons
we include weak magnetism and recoil corrections from
\cite{horowitz:02}. For coherent scattering on heavy nuclei we follow
the implementation from \cite{brt:06}. These rates use the average
nuclear mass ($\bar{A}$) and average nuclear charge ($\bar{Z}$) from
the EOS and include form factor corrections due to decoherence and an
electron polarization correction. They also include ion-ion
correlations from \cite{horowitz:97}. Elastic scattering off 
$\alpha$ particles uses the heavy nuclei scattering cross section with
$A=4, Z=2$, but drops the corrections.

{\emph{Inelastic Neutrino-Electron Scattering:}} \code{NuLib}
calculates the zeroth and first moment of the neutrino-electron
scattering kernel as computed in \cite{bruenn:85}. These kernels are
computed as a function of electron chemical potential (or
specifically, $\eta_e = \mu_e/T$), temperature, and both incoming and
outgoing neutrino energy. In \code{GR1D}, after interpolation, we
enforce the in/out symmetry of the scattering kernels
$R_\ell^\mathrm{in}(\epsilon^\prime,\epsilon) =
R_\ell^\mathrm{out}(\epsilon,\epsilon^\prime)$ as well as
$R_\ell^\mathrm{out}(\epsilon,\epsilon^\prime) =
\exp\left[-(\epsilon^\prime - \epsilon)/T\right] \times
R_\ell^\mathrm{out}(\epsilon^\prime,\epsilon)$
\citep{cernohorsky:94b}. This has the added benefit of reducing the
number of interpolations needed.

This set of interactions is motivated by and is equivalent to the set
used in \cite{liebendoerfer:05}, with the exception that \code{NuLib}
includes weak-magnetism and recoil corrections. In the CCSN test
problems that follow in \sref{sec:s15compare} we use this set of
interactions, ignoring the weak-magnetism and recoil corrections for
the sake of comparison. This set is in no way complete, or even
modern.  More complete neutrino interaction rates are available in the
literature, most notably inelastic neutrino-nucleon scattering
\citep{reddy:98}, and electron capture rates on heavy nuclei
\citep{langanke:03}.

\section{Radiation Test Problems}
\label{sec:tests}

In this section we rigorously test our transport implementation against
classic radiation test problems. We begin with several classical
radiating spheres in \sref{sec:radiatingspheres} and move on to
gravitating radiating spheres with a background fluid motion in
\sref{sec:gravitatingspheres}. In \sref{sec:diffusion}, we further
test our explicit flux treatment by performing a diffusion wave test.

\subsection{Classical Radiation Spheres}
\label{sec:radiatingspheres}

\begin{figure*}[ht]
\centering
\includegraphics[width=0.8\columnwidth]{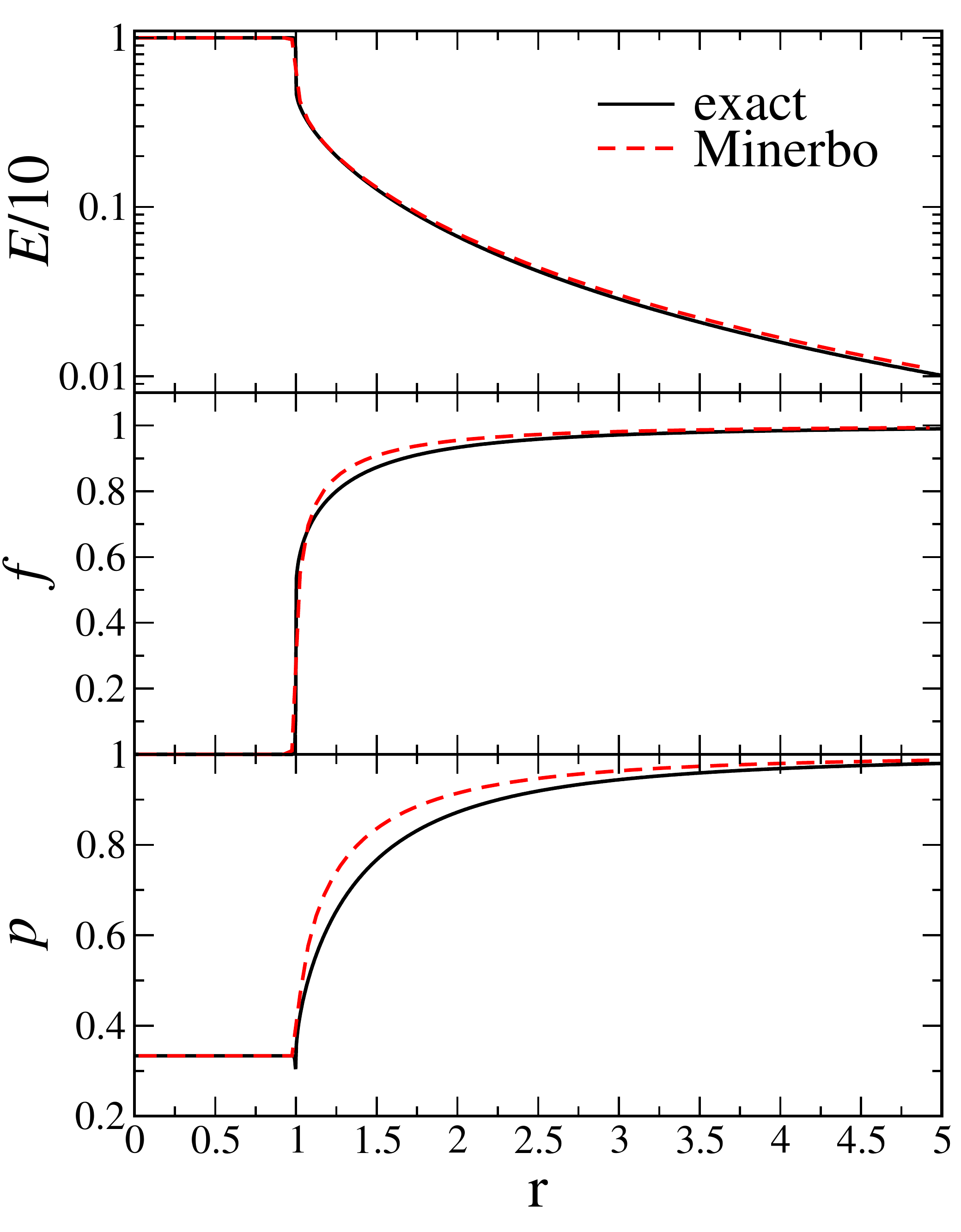}\ \ \ \ \ \
\ \
\includegraphics[width=0.8\columnwidth]{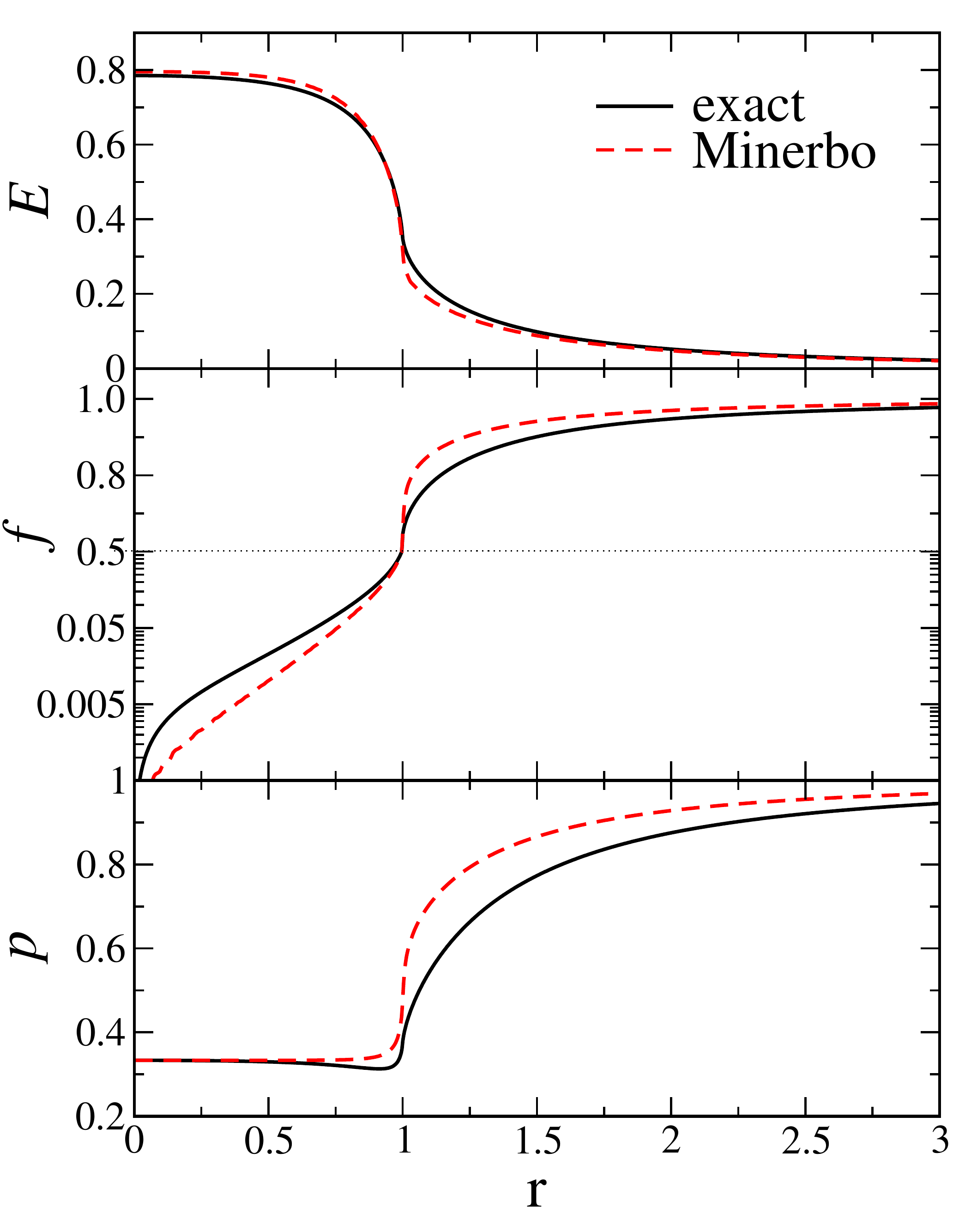}
\caption{Classical radiating spheres with two different neutrino
interaction rates for the interior of the sphere. In the left panel we
repeat the test of \cite{abdikamalov:12} which is a highly opaque sphere with
an optical depth at the center of 250. In the right panel we repeat
the test of \cite{smit:97} that models a sphere with a much lower
opacity, here the optical depth in the center is 4. In both panels we
show, from the top down, the energy density ($E$), the flux factor
($f=F_r/E$), and the Eddington factor ($p=p_{rr}E$). The solid black
line denotes the exact solution while the dashed red line is our
numerical result using the Minerbo closure. We note the vertical axis
in the right-middle panel transitions at the dotted line from linear
for $f>0.5$ to logarithmic for $f<0.5$, this is to clearly show
the behavior in both regimes.}\label{fig:radiatingspheres}
\end{figure*}

We perform two classical radiating sphere tests. In both tests the
sphere has a radius of 1 and there is no scattering opacity. The first
test is one where the absorption opacity is chosen such that the
optical depth in the center of the sphere is very large, $\tau = 250$,
with a very coarse grid (100 evenly spaced zones out to a radius of 5)
such that the Peclet number is Pe=12.5 for cells interior to the
radiating sphere. The corrections we impose on the explicit fluxes
dominate for this test. The value of $a$ in \eref{eq:peclet} is
$\sim$0.08 for zones inside the sphere and 1 outside. The matter
interaction variables follow from \cite{abdikamalov:12} ($b=10$;
$\kappa_a=250$; $r_\mathrm{surface} = 1$). The second radiating sphere
is more optically thin. The opacity is chosen such that the optical
depth in the center of the sphere is 4. This test uses a very fine
grid (800 evenly spaced zones out to a radius of 3) such that the
Peclet number for interior zones is Pe=0.015. The corrections are
non-existent, $a$ in this case is 1 everywhere. The matter interaction
variables follow from \cite{smit:97} ($b=0.8$; $\kappa_a=4$,
$r_\mathrm{surface}=1$). For both cases, analytic solutions exist (see
\citealt{smit:97}). In \fref{fig:radiatingspheres} we show the
results of these two tests using \code{GR1D} (dashed lines) along with
the analytic result (solid lines). The high opacity test results are
shown in the left panel and the low opacity test results are shown in
the right panel. We show the energy density ($E$), flux factor
($f=F_r/E$), and the Eddington factor ($p=p_{rr}/E$). These tests are
Newtonian--there are no general relativistic or velocity effects. Our
transport scheme does well in both the optically thick (with both high
and low Peclet number) region and in the free streaming region. The
results of this test are particularly sensitive to the closure
relation. For our results we use the Minerbo closure. \cite{smit:97}
found significant variation of their results with the choice of
closure. The variation is of order what we see in
\fref{fig:radiatingspheres}. The optimal choice of closure is an
outstanding question, but will not be addressed here.

Since we use an evenly-spaced grid, and all of the zones outside of
the sphere are free streaming, the first-order explicit flux
calculation with a Courant factor of 0.5 gives artifacts in the
radiation moments exterior to the sphere. For this test we avoid this
by evolving the spatial flux in low Peclet number cells
($\mathrm{Pe} < 0.01$) via the second-order explicit midpoint
method. The ultimate solution is the implementation of higher order
implicit-explicit methods valid in the optically thick regime. This is
beyond the current work, but as noted in \sref{sec:exp_flux}, in
typical CCSN conditions and with the grid we use in \code{GR1D} we
typically do not numerical problems when using the first-order
explicit method.

\subsection{General Relativistic Radiating Sphere with Velocity Field}
\label{sec:gravitatingspheres}

To test the energy coupling terms discussed in \sref{sec:encoupling}
we repeat the tests performed in \cite{mueller:10}. These tests use a
gravitating radiating sphere as a source of a spectrum of neutrinos
which radiate through a velocity field similar to what is found in
CCSNe. For the radiation, we choose a thermal Fermi spectrum with a
temperature of 5\,MeV and zero chemical potential.  We use our
standard energy group spacing. We logarithmically space 18 groups with
the first group centered at 1\,MeV with a width of 2\,MeV.  The
largest bin is centered at $\sim280.5\,$MeV with a width of
$\sim 61\,$MeV. Our spatial grid is also logarithmic with a central
zone spacing of $10^4$\,cm, extending to a radius of $10^9$\,cm over
300 zones. The spatial zone spacing at the velocity feature is the
same as \cite{mueller:10}, $\sim 4\,$km. We make the absorption
opacity in the interior of the radiating sphere sufficiently high such
that, to a good approximation, all of the escaping radiation comes
from the surface (like the first radiating sphere test case above). A
velocity profile similar to what is found in the stalled shock phase
of CCSNe is also used. These two features test both the $R_\alpha$
(gravitating) and $O_\alpha$ (accelerating) momentum flux terms. We
test three cases: \emph{a)} a 9.89\,km radiating sphere of negligible
mass with the velocity profile used in \cite{mueller:10}, \emph{b)} a
9.89\,km radiating sphere of dust with a constant density of
$9\times10^{14}$\,g\,cm$^{-3}$ such that the gravitating mass is
$\sim 1.833\,M_\odot$ but no velocity field, \emph{c)} both the
gravitating radiating sphere and the velocity profile used in the
previous cases.

The velocity profile is taken from \cite{mueller:10},
\begin{equation}
v(r) = \begin{cases} 0; & r \le  135\,\mathrm{km} \\
  -0.2c\frac{r-135\,\mathrm{km}}{15\,\mathrm{km}}; & 135\,\mathrm{km} < r < 150\,\mathrm{km} \\
  -0.2c\left(\frac{150\,\mathrm{km}}{r}\right)^2; & r \ge 150\,\mathrm{km}  \end{cases}\,.\label{eq:velocityprofile}
\end{equation}

We will first discuss the analytic solutions which also follow closely
to those in \cite{mueller:10}. For the luminosity, the analytic
expression can be determined by taking the free streaming limit of
the fluid frame flux (\eref{eq:hr} with $E = F_r/X = P_{rr}/X^2$) and
noting that the energy integrated, static solution of \eref{eq:e}, in
the vacuum limit, gives $\alpha r^2 \int_\epsilon F_r d\epsilon /X^2 =
\mathrm{const}$, where $\int_\epsilon d\epsilon$ denotes the integral
over the energy spectrum. Therefore,
\begin{equation}
\mathcal{L}(r)  = (4\pi r)^2 \int_\epsilon \mathcal{H}^r d\epsilon = (4\pi)^2 
\frac{W}{\alpha}\frac{1-v}{1+v} \frac{\alpha r^2 \int_\epsilon F_r d\epsilon}{X^2}\,. \label{eq:l}
\end{equation}
It follows from this that in the free streaming limit, variations in
the fluid frame luminosity should arise only due to non-zero velocities
and gravitational redshift,
\begin{equation}
\mathcal{L}(r)  \propto \frac{W}{\alpha}\frac{1-v}{1+v}\,.
\end{equation}

\begin{figure*}[ht]
\centering
\includegraphics[width=0.94\columnwidth]{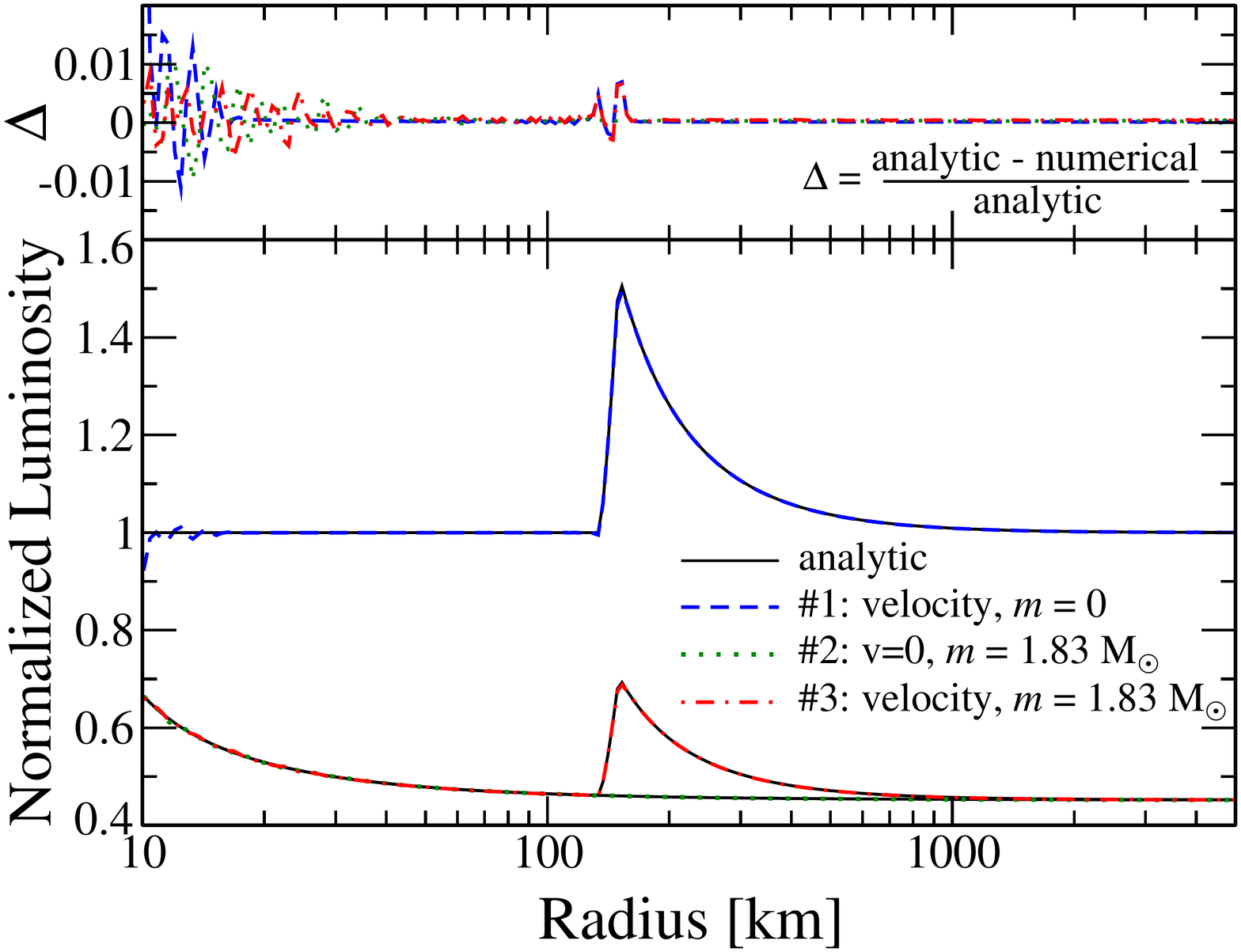}\ \
\ \ \ \ \ \ 
\includegraphics[width=0.94\columnwidth]{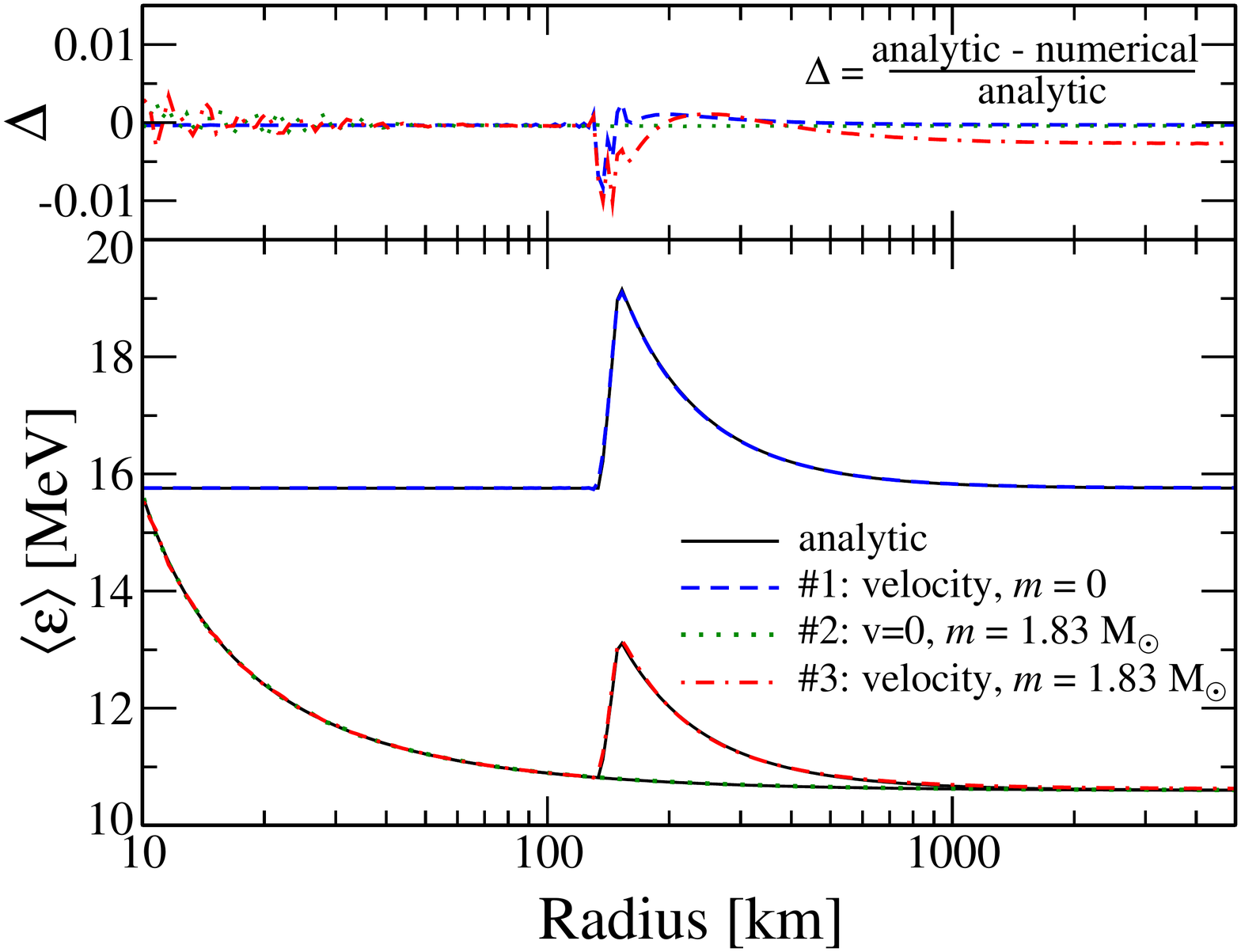}
\caption{Gravitating radiating spheres with a velocity field based on
the test cases of \cite{mueller:10}. We conduct three separate
radiating sphere simulations: the first sphere (\#1) has a negligible
mass but has the velocity profile of \eref{eq:velocityprofile}, the
second test (\#2) has no velocity field, but the sphere has a
gravitational mass of 1.833\,$M_\odot$, the third case (\#3) is both
gravitating and contains the velocity profile of
\eref{eq:velocityprofile}. In the left figure we show the fluid frame
luminosity, normalized to the sphere's value at infinity. We also
apply the expected factor of $\alpha_\mathrm{surface}^2 \sim 0.452$ to
this normalization. In the right figure we show the fluid frame
average energy. For both figures we include the analytic prediction,
and also the relative deviation ($\Delta$) of our simulations from
these analytic predictions.}\label{fig:muellertests}
\end{figure*}

The standard relativistic Doppler effect applies for the average
energy as measured in the fluid frame with the general relativistic
addition here to account for the gravitational redshift as the
neutrinos stream away from the source,
\begin{equation}
\langle \epsilon \rangle(r) =
\frac{\alpha(r_\mathrm{surface})}{\alpha(r)}
\frac{\langle \epsilon_\mathrm{surface} \rangle }{W(1+v)} \sim \frac{\alpha(r_\mathrm{surface})}{\alpha(r)}
\frac{15.7568\,\mathrm{MeV}}{W(1+v)}\,,\label{eq:analen}
\end{equation} 
where the 15.7568\,MeV is the average neutrino energy of a Fermi
distribution with zero chemical potential and a temperature of 5\,MeV.
The analytic solution motivates the high opacity mentioned above so
that all of the radiation that reaches infinity was originally emitted
from $r=r_\mathrm{surface}$ and not from deeper in where the
appropriate redshift correction factor would be different than
$\alpha(r_\mathrm{surface})$. For reference,
$\alpha(r_\mathrm{surface}) = 0.6723$.

In \fref{fig:muellertests} we show the fluid frame neutrino luminosity
and average energy as a function of radius from \code{GR1D} for each
of these three tests. We also show analytic solutions as solid lines.
For the luminosities, we normalize the solution to the simulation
value at $r=10^4\,$km and then scale every solution by
$\alpha_\mathrm{surface}^2$. For the gravitating spheres, in addition
to the redshift effect, the luminosity is suppressed by an additional
factor of $\alpha_\mathrm{surface}$ due to general relativistic time
dilation at the source. For the average energy analytic solution, we
use \eref{eq:analen} directly and do not normalize the data in any
way.  The data from \code{GR1D} are shown as the blue dashed lines
(case \#1), green dotted lines (case \#2), and red dotted-dashed lines
(case \#3). Additionally, we show the relative difference between the
analytic solution and our data. The observed deviations are $<1\%$ for
both the luminosity and the average energy. The effect of the first
order explicit flux calculation can be seen for small radii. To be
clear, since we evolve the lab frame variables $E$ and $F_r$ and
compute the fluid frame value shown here with \eref{eq:hr} and the
closure, this test shows that the total luminosity in the lab frame is
constant as a function of radius (to better than 1\% at $r=150\,$km),
as expected. However, even though the total luminosity in the lab
frame is constant, this does not mean the distribution of the energy
among the energy groups remains constant. Recall that our definition
of neutrino energy applies only in the fluid rest frame, therefore
when fluid velocities exist, even if the neutrinos are completely
decoupled from the fluid ($\eta = \kappa_a = \kappa_s = 0$) the energy
of a particular energy group in the lab frame is not the same as the
fluid frame. When the fluid velocity is changing (either in time or
radially), this induces a flux between neighboring energy groups which
is captured in $O_\alpha$.

These tests provide a clean environment to test for neutrino number
and energy conservation in our code.  By tracking the energy and
neutrino number emitted/absorbed by the matter in the radiating sphere
plus the energy and neutrino number exiting the outer boundary and
computing the energy and neutrino number present on the grid we can
track violations.  For test case \#1, \#2, and \#3, the neutrino
number violation is $\sim$0.5\%, $\sim$2\%, and $\sim$2\% respectively
while the energy conservation is $\sim$0.5\%, $\sim$0.03\%, and
$\sim$0.07\%, respectively.  For a test case with neither velocity nor
a gravitating mass, the number violation and energy violation is $\sim$0.01\%.
The majority of the lepton violation in the gravitating sphere cases
(\#2 and \#3) occurs in the sphere itself rather than the free
streaming region outside.

\subsection{Diffusion Wave}
\label{sec:diffusion}

The final basic code test we perform is a diffusion wave test. This
test is identical to the diffusion test of \cite{pons:00}. The purpose
of this test is to show the ability of our explicit flux
implementation to perform in very diffusive conditions that would
normally fail without the corrections made in \sref{sec:exp_flux}.
The test problem is to follow the diffusion of a Dirac delta function
of radiation located at the origin at $t=0$. In the diffusion limit
the analytic solution is,
\begin{equation}
E = \left(\frac{\kappa}{t}\right)^{3/2}\exp\left(\frac{-3\kappa
    r^2}{4t}\right);\ \ \ \ \ F_r = \frac{r}{2t}E\,.
\end{equation}
Following \cite{pons:00}, we take a spherical grid extending to a
radius of 1 using 100 equally spaced zones. We do two tests on this
grid. Test A takes a scattering opacity $\kappa_s=100$ (giving a
$\mathrm{Pe}=1$) and starts the simulation from $t=1$ to avoid the
delta function at $t=0$. Test B takes a scattering opacity of
$\kappa_s=10^5$ ($\mathrm{Pe}=1000$) and starts the simulation at
$t=200$. We sample our simulations at three additional times, $t=2,3,$
and 5 for test A and $t=240,300,$ and 400 for test B. Unlike
\cite{pons:00} we use our standard closure shown in \eref{eq:closure}
rather than taking $p_{rr} = 1/3$, however, we find such a choice
makes little difference since the momentum density is always much less
than the energy density.

\begin{figure*}[ht]
\centering
\includegraphics[width=1.98\columnwidth]{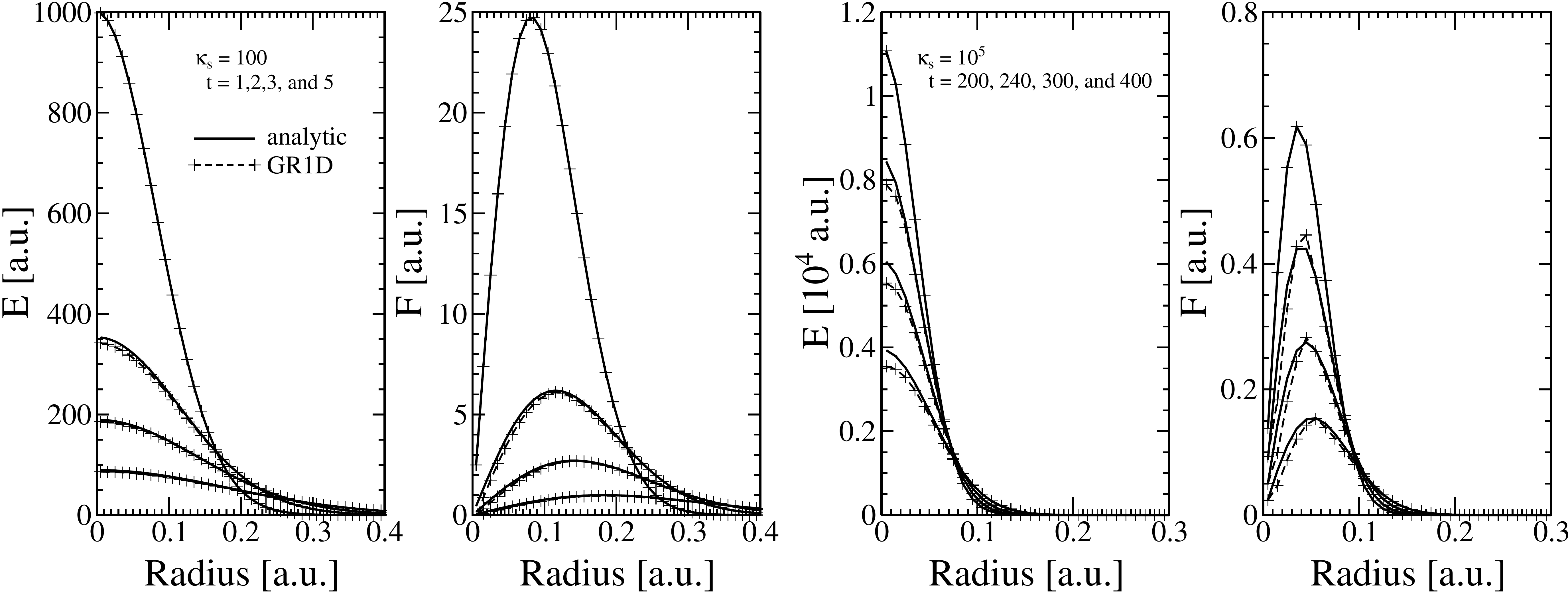}
\caption{Diffusion of a centralized packet of radiation. In the left
  two panels are the results of a test where the grid spacing and
  opacity give a Peclet number of 1. The simulations depicted in the
  right two panels however use a much larger opacity and the Peclet
  number is 1000. The left panel of each simulation shows the energy
  density, the right panel shows the momentum density. As time
  progresses and the radiation diffuses out, the curves decrease in
  magnitude. The tests are identical to those in
  \cite{pons:00}.}\label{fig:diffusionwave}
\end{figure*}

In \fref{fig:diffusionwave} we show our results compared to the
analytic solution. In the left two panels are the results of test A
with the energy density on the left and momentum density on the right.
In the right two panels are the results of test B, again with energy
density on the left and momentum density on the right. We note the
first result in both cases ($t=1$ and $t=200$ in test A and B,
respectively) is the initial setup, and therefore the \code{GR1D}
result is identical to the analytic result. \code{GR1D} can reproduce
the diffusion limit analytic solution quite well. For test B, we do
find differences that we can explain in part from our interpolation
choice between the diffusion limit and hyperbolic limit of the flux
determination (i.e. the value of $a$ in \eref{eq:energyflux} and
\eref{eq:momentumflux}). If we assume $a=0$ in order to force the
spatial flux calculation to be determined by the diffusion
approximation we can achieve a result closer to the analytic solution
than what is shown in \fref{fig:diffusionwave}. We are not worried
about the differences seen here significantly affecting our CCSN
simulations for several reasons.  First, a Peclet number of 1000 is
quite large.  In CCSN simulations we only reach these values for the
largest energy groups ($\sim > 150$\,MeV) in high density zones. Also,
the simulation times we consider are much less than the diffusion time
of the neutrinos in regions with these high Peclet numbers.  For
applications of this code to protoneutron star cooling it will be
necessary to further explore this regime.

\section{Core-Collapse Supernovae Test Problems}
\label{sec:ccsntests}
\subsection{Comparison with other Neutrino Transport Codes}
\label{sec:s15compare}

A somewhat standardized test of spherically symmetric, general
relativistic, neutrino radiation hydrodynamics is the collapse,
bounce, and early post-bounce evolution of the 15$\,M_\odot$
progenitor star of \cite{ww:95} using the EOS from \cite{lseos:91}
with $K_0=180\,$MeV (LS180) and a set of simplified neutrino rates
from \cite{bruenn:85}. The LS180 EOS has a maximum cold neutron star
gravitational mass of 1.84\,$M_\odot$. This maximum mass has been
ruled out by the observation of $\sim$2\,$M_\odot$ neutron stars
\citep{demorest:10,antoniadis:13}, however, it has been used in
several other studies as a basis for comparison
\citep{liebendoerfer:05,mueller:10}. For this reason alone, we also
use it here. The simulation data of \cite{liebendoerfer:05} are
publicly available via the publisher's website. We use that data here
to show \code{GR1D}'s ability to simulate the core collapse, bounce,
and post-bounce phases of a CCSN in spherical symmetry. It is worth
mentioning that the weak magnetism and recoil corrections of
\cite{horowitz:02} are not included in the work of
\cite{liebendoerfer:05}, and therefore are also not included in this
comparison. The comparison data in \cite{liebendoerfer:05} is between
two simulation codes, \emph{Agile}-\code{BOLTZTRAN} and \code{VERTEX}.
\emph{Agile}-\code{BOLTZTRAN} is a fully general relativistic,
Lagrangian hydrodynamics code with a Boltzmann neutrino transport
solver while \code{VERTEX} is a Newtonian, Eulerian hydrodynamics code
with a gravitational potential correction to mimic the effects of
general relativity. \code{VERTEX}'s transport is a moment scheme with
a variable Eddington factor solved via a model Boltzmann equation.
Improvements to \code{VERTEX} after \cite{liebendoerfer:05}, and other
comparisons \citep{marek:06,mueller:10}, suggest that the publicly
available data in \cite{liebendoerfer:05} from
\emph{Agile}-\code{BOLTZTRAN} are more reliable than those from
\code{VERTEX}.

Our simulation uses the following energy grid for each neutrino type,
we logarithmically space 18 groups with the first group centered at
1\,MeV with a width of 2\,MeV. The largest bin is centered at
$\sim280.5\,$MeV with a width of $\sim 61\,$MeV. We assume three
neutrino species by lumping $\nu_\mu$, $\bar{\nu}_\mu$, $\nu_\tau$,
and $\bar{\nu}_\tau$ into a characteristic heavy-lepton neutrino
$\nu_x$. The spatial grid is a hybrid grid where the inner 20\,km
is evenly spaced with 100 zones of 200\,m each. Outside of this radius
we use a logarithmically spaced grid consisting of 550 zones from
200\,km out to $\sim$15000\,km. To be clear, for the following comparison
simulation we use an explicit treatment of inelastic neutrino-electron
scattering and the pair-production approximation discussed in
\sref{sec:nusourceterms} and \sref{sec:nulib}. We explore these
approximations in the following section. 

\begin{figure}[t]
\centering
\includegraphics[width=\columnwidth]{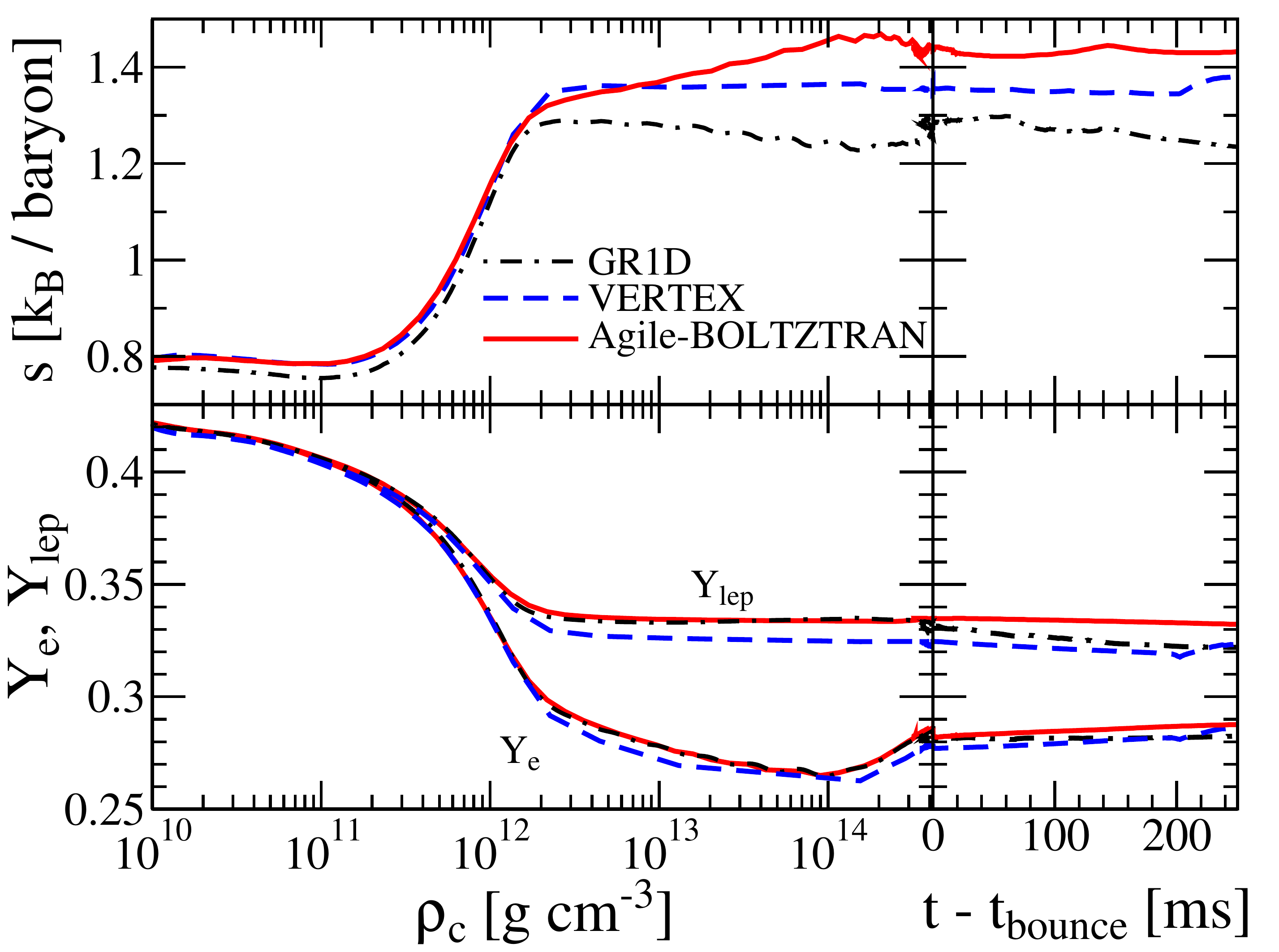}
\caption{Evolution of central entropy and lepton numbers. In the left
(right) panels are the pre-bounce (post-bounce) values of the central
entropy [top] and electron ($Y_e$) and total lepton ($Y_e + Y_\nu$)
fraction [bottom]. Pre-bounce we show these values versus the central
density, while after bounce we show them versus time. \code{GR1D}'s
values are the black dashed-dotted lines, \code{AGILE} results are
shown as red solid lines and the \code{VERTEX} results are shown as
blue dashed line.}\label{fig:central_yeandentropyvsrhotime}
\end{figure}

In \fref{fig:central_yeandentropyvsrhotime}, we show the evolution of
the central entropy (top panels) and the electron ($Y_e$) and total
lepton number ($Y_\mathrm{lep} = Y_e + Y_\nu$) fractions (bottom
panels) along with the \emph{Agile}-\code{BOLTZTRAN} and \code{VERTEX}
results from \cite{liebendoerfer:05}. To more clearly show the
evolutionary changes we split the simulation into the pre-bounce phase
(left) and post-bounce phase (right). Most of the interesting neutrino
physics happens in the final $\sim$10-20\,ms of the collapse phase as the
central density rises from $\sim 10^{11}$g\,cm\,$^{-3}$ to
$\sim 10^{14}$\,g\,cm$^{-3}$, therefore we show this phase versus central
density. After bounce we plot the quantities versus post-bounce time
as the central density during this time is essentially constant. We
first discuss the lepton fraction. The agreement between the
\code{GR1D} and the Boltzmann solutions, particularly with
\emph{Agile}-\code{BOLTZTRAN}, during the collapse phase is
exceptional and relies heavily on the implementation of inelastic
neutrino-electron scattering, energy bin coupling, and neutrino
advection with the fluid in the optically thick medium. However, we
note that \cite{lentz:12b} have shown that improved electron capture
rates on heavy nuclei beyond what was used in \cite{liebendoerfer:05}
can play the role that inelastic neutrino-electron scattering plays
here. There are small oscillations in both the neutrino and electron
fraction between densities around $10^{13}$\,g\,cm$^{-3}$ and nuclear
density that are the result of equilibration between the electrons and
neutrinos as the electron chemical potential of the matter rises and
the discretely spaced neutrino energy levels fill up. This is present
in both the \emph{Agile}-\code{BOLTZTRAN} results and ours and can be
eliminated with increased energy resolution as has been shown by
\cite{rampp:02}. The \code{VERTEX} data from \cite{liebendoerfer:05}
has too low of temporal resolution to see this effect. The total
lepton fraction is essentially constant after the electron neutrinos
become fully trapped, which occurs in the core, around
$\rho\sim 2\times 10^{12}$\,g\,cm$^{-3}$. For the post-bounce
evolution, over the simulated 250\,ms, the central lepton and electron
fractions remain roughly constant with some secular drift. We find
that the central neutrino fraction drops similarly in all three codes,
but that \code{GR1D}'s central electron fraction does not increase as
is seen in the Boltzmann codes, but rather remains constant.

\begin{figure*}[t]
\centering 
\includegraphics[trim=0 3cm 0 0, clip, width=1.98\columnwidth]{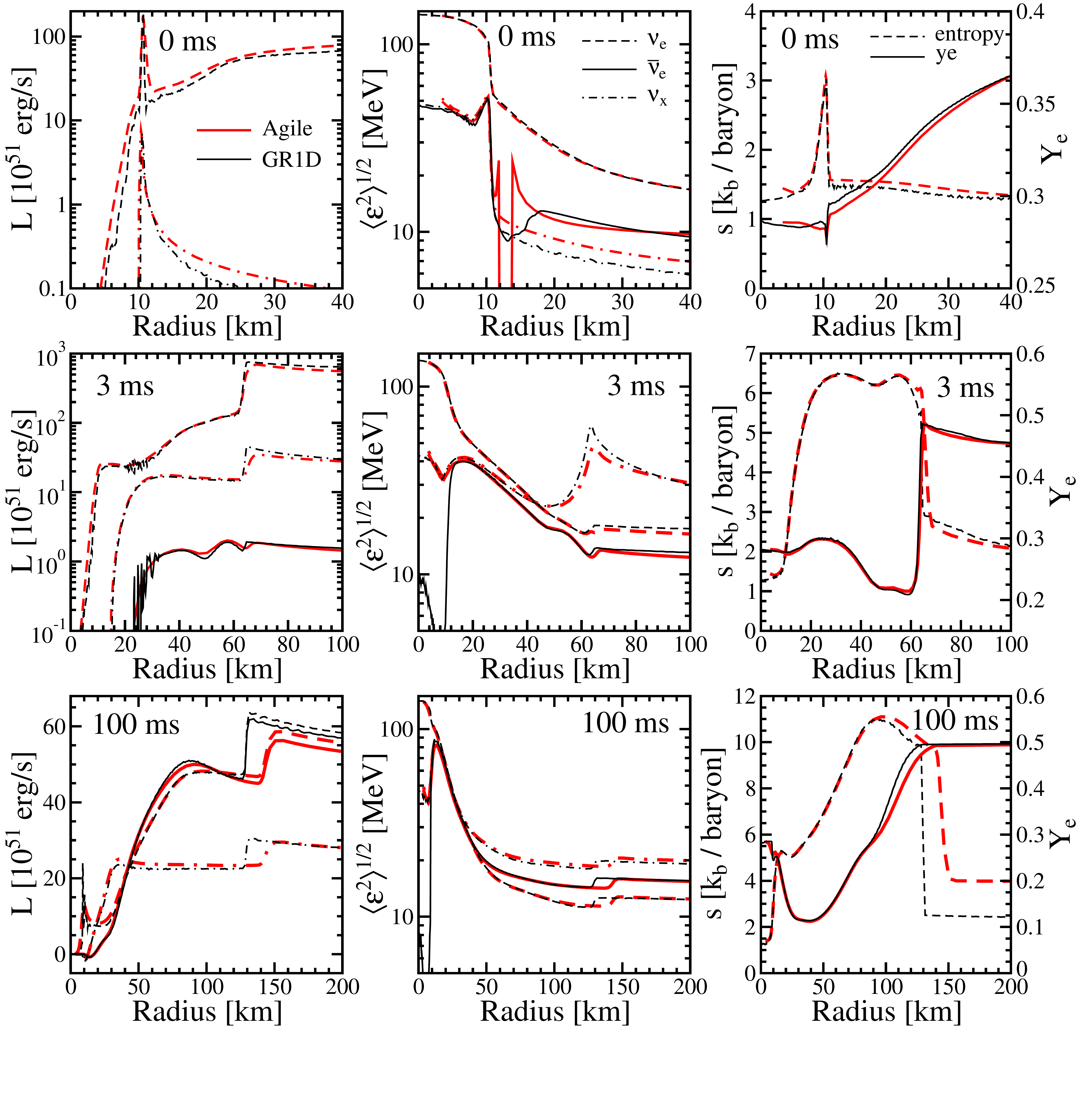}
\caption{Comparison of radial profiles. We show various radial
profiles from \code{GR1D} (thinner black lines) and
\emph{Agile}-\code{BOLTZTRAN} (thicker red lines) at three times:
bounce (top panels), 3\,ms after bounce (middle panels), and 100\,ms
after bounce (bottom panels). At each time we show each neutrino
flavor's luminosity (left panels), root mean squared energy (center
panels), and profiles of the entropy and electron fraction (right
panels). For the neutrino quantities, the profiles for electron
neutrinos use a dashed line, electron antineutrinos are shown with a
solid line, and the heavy-lepton neutrino profiles are shown with a
dashed-dotted line. The entropy profiles use a dashed line while the
electron fraction profile lines are solid. Overall, the agreement
between \code{GR1D} and \emph{Agile}-\code{BOLTZTRAN} is excellent.}\label{fig:profiles}
\end{figure*}

In the top panel of \fref{fig:central_yeandentropyvsrhotime} we show
the evolution of the matter entropy in the innermost zone. We again
show the pre-bounce evolution versus central density on the left and
the post-bounce evolution versus time on the right. The different
starting values of the entropy between the codes is likely due to the
different implementations of the nuclear EOS. Aside from this shift,
the evolution is similar. Before trapping, the entropy rises due to
neutrino interactions, but after the onset of neutrino trapping the
matter entropy should remain constant, we see an initially lower value
and then a small decrease as the matter density increases to nuclear
densities. We note that the entropy shown here is only the matter
entropy, it does not include the entropy of the neutrinos, which is
small and roughly constant throughout the trapping region. After
bounce the entropy remains roughly constant, but decreases by about
4\% over the 250\,ms. While concerning, and clearly an area for future
improvement, these aspects of the central core evolution do not have a
large impact on the rest of the protoneutron star within the first few
100\,ms or even up to a second after core bounce since the neutrino
diffusion time is long compared to the times simulated here.

In \fref{fig:profiles} we show a collection of profiles showing the
results of both \code{GR1D} and \emph{Agile}-\code{BOLTZTRAN}. The
\code{VERTEX} results, after improving the pseudo-relativistic
potential in \cite{marek:06}, agree very well with the
\emph{Agile}-\code{BOLTZTRAN} results. We include profiles of the
neutrino luminosity (left panels), neutrino root mean squared energy
(center panels), and electron fraction and entropy (right panels). The
luminosity follows from the first expression in \eref{eq:l} and the
root mean squared energy is computed by averaging $\epsilon^2$ over
the fluid frame neutrino number distribution, equivalent to the
definition in \cite{liebendoerfer:05}. We show these profiles at
bounce (top panels), 3\,ms after bounce (middle panels), and 100\,ms
after bounce (bottom panels). For the heavy lepton neutrino
luminosity, we average the neutrino and antineutrino results from
\emph{Agile}-\code{BOLTZTRAN}. \code{GR1D} agrees very well with the
\emph{Agile}-\code{BOLTZTRAN} results in almost every quantity.

At bounce, defined as when the entropy in the core region first
reaches a value of 3 $k_B/$baryon, the electron antineutrino
production is still highly suppressed from the high electron chemical
potential, its luminosity is off the bottom of the panel. The other
neutrino luminosities have a strong peak in production at $\sim$10\,km
which corresponds to the shock formation radius. We note that the
baryonic mass enclosed in the shock at this time is
$\sim$0.54\,$M_\odot$ in \emph{Agile}-\code{BOLTZTRAN} and
$\sim$0.55\,$M_\odot$ in \code{GR1D}. There are some artifacts in the
root mean squared energy of the electron antineutrino profile at this
time, but the reader is reminded that the total energy in these
neutrinos is very small. The entropy and electron fraction profiles
agree very well at this time. There is a slight difference in the
$Y_e$ of the accreting material, but it is important to note that this
is a very dynamic time. For example, a \code{GR1D} profile from a mere
50\,$\mu$s earlier reproduces the $Y_e$ profile in the accretion
region (outside the shocked core) of the \emph{Agile}-\code{BOLTZTRAN}
profile.

At 3\,ms after bounce, all of the quantities plotted in the middle
panels of \fref{fig:profiles} show exceptional agreement between
\code{GR1D} and \emph{Agile}-\code{BOLTZTRAN}. \code{GR1D} reproduces
every divot, bump, peak, and trough in the luminosity, root mean
squared energy, entropy and electron fraction. The largest discrepancy
is the electron antineutrino root mean squared energy in the core.
However, there is very little total electron antineutrino energy
density in the core, so this is not a large concern. The likely
culprit is our lack of pair production processes (and nucleon-nucleon
Bremsstrahlung in particular which dominates the rate in the
unshocked, dense core) in the electron neutrino/antineutrino sector. A
similar electron antineutrino root mean squared energy was seen in the
Newtonian results of \cite{liebendoerfer:05} where nucleon-nucleon
Bremsstrahlung was omitted. However, even if this was included, the
extreme electron degeneracy in the core keeps the energy density of
electron antineutrinos orders of magnitude lower than that of the
electron neutrinos.

At 100\,ms after bounce, the bottom panels of \fref{fig:profiles} show
that inside of $\sim$100\,km, we continue to see excellent agreement.
In the inner 20-25\,km, the luminosity in all three flavors matches
between the codes, including the inward diffusion of the electron
antineutrino and the heavy-lepton flavor neutrinos near the location
of shock formation and the boundary between the shocked and unshocked
core. The electron neutrino and antineutrino luminosities and energies
agree well between the codes, including the location and strength of
the gain regions, out to $\sim$130\,km, where we reach the shock front
in \code{GR1D}. This is the single biggest difference
between our simulations and those of
\cite{liebendoerfer:05,marek:06,mueller:10}. We discuss this in the
following section when we look at the time evolution of the neutrino
observables. The heavy-lepton neutrino luminosity and root mean
squared energy deviates from the \emph{Agile}-\code{BOLTZTRAN} near
its neutrinosphere. Considering the simplicity of our approximation
for the heavy-lepton neutrino production/annihilation we achieve
remarkable agreement. We explore this in much greater detail in the
following section where we assess our approximations. At the shock
front, since we are showing the fluid frame neutrino quantities, the
luminosity and root mean squared energy jump. Due to \code{GR1D}'s
hydrodynamics, the shock is better resolved, resulting in a sharper
(and larger) jump in the fluid frame luminosities and energies.
Outside of the shock, \emph{Agile}-\code{BOLTZTRAN} burns silicon to
nuclear statistical equilibrium (NSE) when the temperature surpasses
0.44\,MeV, whereas in \code{GR1D} we incorrectly, but for simplicity,
assume NSE everywhere. This results in differences in the entropy
outside of the shock.

\begin{figure}[t]
\centering
\includegraphics[width=\columnwidth]{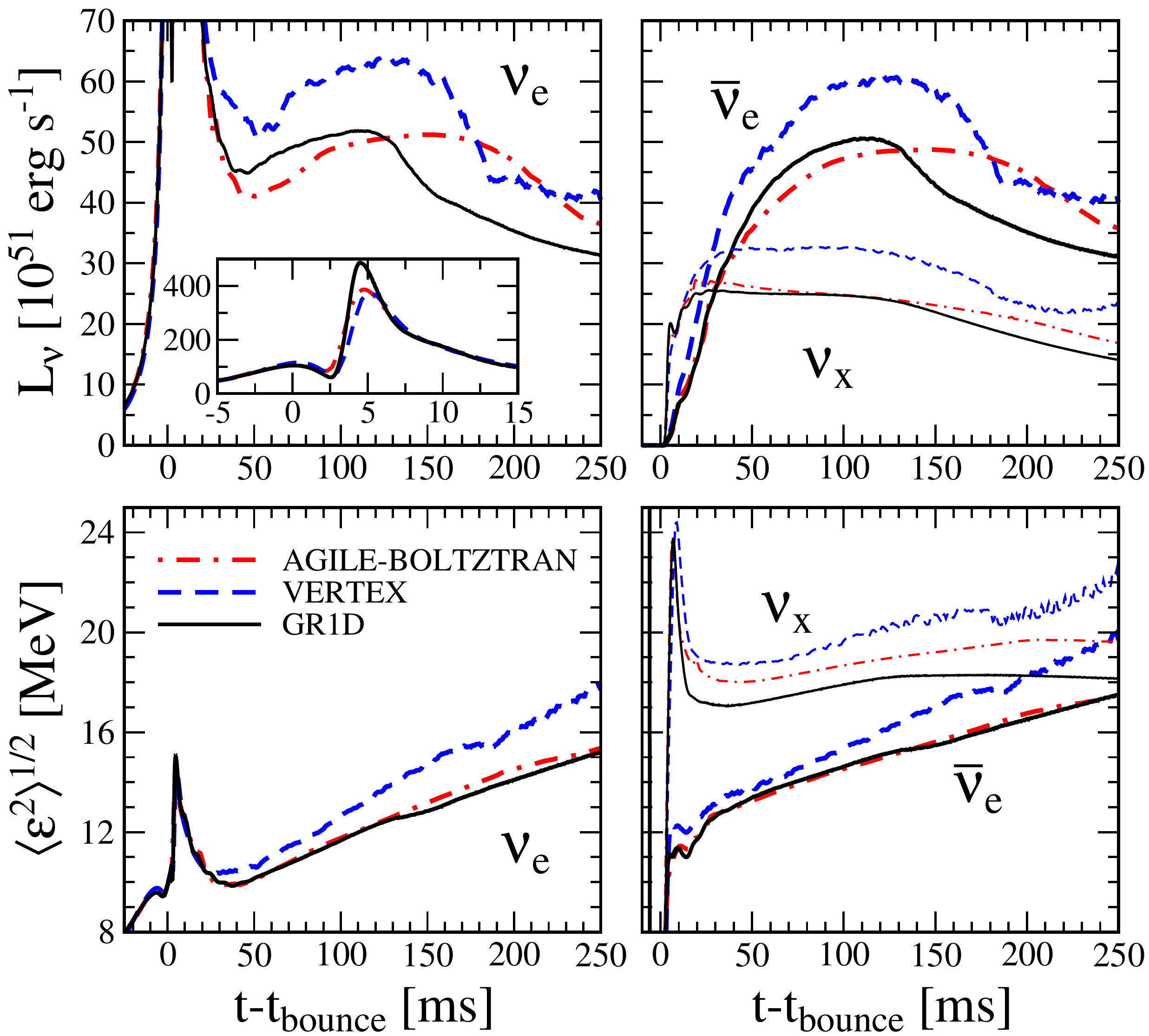}
\caption{Comparison of time evolutions of key neutrino quantities. We
show the time evolution of the neutrino luminosity (top panels) and
the neutrino root mean squared energy (bottom panels) as measured by
an observer moving with the fluid at 500\,km. Results of \code{GR1D}
are shown as solid black lines, \emph{Agile}-\code{BOLTZTRAN} results
are shown as dashed-dotted red lines, and \code{VERTEX} results are
shown as dashed blue lines. The left two panels are the luminosity and
root mean squared energy of electron neutrinos, the right two panels
show this information for both electron antineutrinos (thick lines)
and the heavy-lepton neutrino (thin lines). The inset
in the upper left panel show just the deleptonization burst near the
time of bounce.}\label{fig:lumsandenergies}
\end{figure}

As another comparison, we look at the far-field neutrino luminosities
and root mean squared energies versus time for the first 250\,ms after
bounce. Following the convention of \cite{liebendoerfer:05}, we show
these quantities in the fluid frame at 500\,km. However, we note that
the velocity at this radius can be $\sim$-0.06\,$c$ and therefore
these energies and luminosities are roughly 6\% and 12\% larger,
respectively, than one would observe at Earth. We compare our results
to those obtained with both \code{VERTEX}\footnote{Improvements made
to the \code{VERTEX} code after the publication of
\cite{liebendoerfer:05} change the predicted neutrino luminosities
\citep{marek:06}. These changes lead to lower luminosities that have
magnitudes similar to the \emph{Agile}-\code{BOLTZTRAN} results.
However, and this is relevant to our results, the sharp drop in the
neutrino luminosity around $\sim180\,$ms remains.} and
\emph{Agile}-\code{BOLTZTRAN}. In \code{GR1D}, by default, we do not
include electron antineutrinos or heavy-lepton neutrinos before the
central density reaches $10^{12}$\,g\,cm$^{-3}$ as they have very
little luminosity and no dynamical effect on the simulation. We show
these comparisons in \fref{fig:lumsandenergies}. This plot is very
similar to the version we presented in \cite{oconnor:13}, however, our
transport code has been significantly improved since then. The average
energies from \code{GR1D} over the entire 250\,ms are in excellent
agreement with the \emph{Agile}-\code{BOLTZTRAN} results. The largest
difference is in the heavy-lepton neutrino root mean squared energy
that we under-predict by $\sim$1\,MeV. ($\sim$6\%) compared to the
\emph{Agile}-\code{BOLTZTRAN} simulation. We comment on this further
in \sref{sec:approxs}.

The luminosities between \code{GR1D} and \emph{Agile}-\code{BOLTZTRAN}
up to $\sim$100\,ms are in general agreement. We note that the
luminosities of \cite{mueller:10} are also generally higher than those
of \emph{Agile}-\code{BOLTZTRAN} in the first 100-150\,ms, and also
during the deleptonization peak. This may be due to the higher
resolution at the shock front that we discussed in the context of the
radial luminosity profiles above, an effect which is also seen in the
simulations of \cite{mueller:10}. The heavy-lepton neutrino luminosity
follows the \emph{Agile}-\code{BOLTZTRAN} results very closely. We
discuss this agreement more in the context of exploring the
approximation we make for the pair-production processes in the
following section (\sref{sec:approxs}). At $\sim$130\,ms, the
silicon-oxygen interface accretes through the shock and causes a drop
in the neutrino luminosities. A similar effect is seen in the
\code{VERTEX} data, although at a later time. The adaptive grid of
\emph{Agile}-\code{BOLTZTRAN} smears out the jump in density at the
interface and such a steep drop in neutrino luminosities is not
seen. It is currently unknown why the silicon-oxygen interface
accretes through the shock earlier in \code{GR1D} when compared to
\code{VERTEX}.  However, the difference in the collapse times
($\sim$225\,ms in \code{GR1D} compared to $\sim$170\,ms in
\code{VERTEX} and \emph{Agile}-\code{BOLTZTRAN}) is suggestive that
the problem is rooted in the low density EOS which as mentioned above,
does not include any burning or non-NSE physics. This problem is
particularly difficult to diagnose since core collapse is a critical
phenomenon.  Investigations are ongoing. 

\begin{figure}[t]
\centering
\includegraphics[width=\columnwidth]{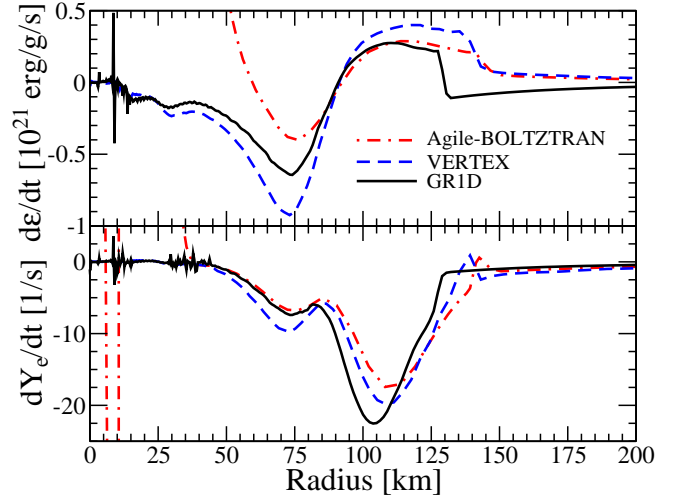}
\caption{Rates of change of specific internal energy (top panel) and
  lepton number (bottom panel) versus radius at a postbounce time of
  100\,ms. In addition to the \code{GR1D} results shown as black solid
lines, we show the results from \cite{liebendoerfer:05}. Red dashed-dotted
lines are the \emph{Agile}-\code{BOLTZTRAN} data, while the blue
dashed lines are the \code{VERTEX} results.}\label{fig:exchangerates}
\end{figure}

A crucial role of neutrinos in CCSNe is to deposit energy in the outer
layers of the postshock region.  This deposition leads to the
generation of the so-called gain region, where there is a net positive
exchange of energy from the radiation field to the matter.  The
efficiency of energy absorption in the gain region is small $(\lesssim
10\%)$, but it is likely a very important aspect of supernova shock
revival.  We look at several quantities to ensure that our transport
scheme is adequately capturing the effects of neutrino heating.  In
\fref{fig:exchangerates}, we show the rate of energy (top) and lepton
(bottom) exchange from the neutrinos to the matter as a function of
radius, at 100\,ms after bounce\footnote{In practice, we change the
  conservative hydrodynamic quantities which are non-linear
  combinations of $\rho$, $P(\epsilon)$, $Y_e$, and $v$.  For this
  figure, we approximate $dY_e/dt$ as \eref{eq:dye} divided by
  $D\Delta t$ (the $D=\rho W X$ quantity is conserved over the
  radiation time step). We show $d\epsilon /dt$ as \eref{eq:dtau}
  divided by $\rho \Delta t$.}. As in the other figures, we show the
\emph{Agile}-\code{BOLTZTRAN} and \code{VERTEX} results as red
dashed-dotted and blue dashed lines, respectively. In all three
simulations, interior to the gain region (where $d\epsilon/dt > 0$)
all simulations show a significant cooling region.  As mentioned
above, the \code{VERTEX} simulation over-predicts the neutrino
luminosity which is a result of the higher cooling rate seen
here. \emph{Agile}-\code{BOLTZTRAN}'s $d\epsilon/dt$ in the public
data suffers from a bookkeeping error which results in the energy and
lepton exchange rates being calculated for output when the electron
neutrino and electron antineutrino are slightly out of equilibrium
\citep{liebendoerfer:15pc}. This makes a comparison at small radii
(and even into the cooling region) difficult.  In all codes, the gain
region begins at $\sim$90\,km and extends out to the shock (which is
$\sim$15-20\,km lower in \code{GR1D}). Outside of the shock
\code{GR1D} returns to a net cooling, while the other simulations
continue to show net heating.  This is likely due to the thermodynamic
conditions outside of the shock. In the Boltzmann transport codes the
entropy is much higher outside the shock from a more complete
treatment of the low density, non-NSE material. This gives a higher
abundance of free neutrons, which increases the opacity for
antineutrino capture in this region. The lepton exchange rates from
\code{GR1D} also track the full Boltzmann results fairly well.  The
main differences are near the shock and originate from the different
shock locations.

\begin{figure}[t]
\centering
\includegraphics[width=\columnwidth]{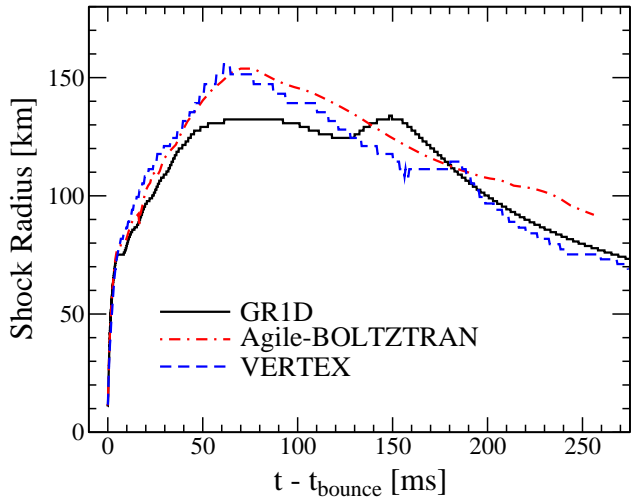}
\caption{Shock radius evolution over time. In addition to the
  \code{GR1D} results shown as black solid lines, we show the results
  from \cite{liebendoerfer:05}. Red dashed-dotted lines are the
  \emph{Agile}-\code{BOLTZTRAN} data, while the blue dashed lines are
  the \code{VERTEX} results.}\label{fig:shock}
\end{figure}

As a final comparison, in \fref{fig:shock}, we show the shock
evolution in \code{GR1D}, \emph{Agile}-\code{BOLTZTRAN}, and
\code{VERTEX}.  The most significant difference is the extent to which
the shock reaches before it ultimately starts to recede, \code{GR1D}
under-predicts this by $\sim$20\,km. As was apparent in the electron
type neutrino luminosity evolution, the time of the silicon-oxygen
interface accretion through the shock can also be seen in the shock
radius evolution in both \code{GR1D} and \code{VERTEX}.  The accretion
of this interface results in a brief period of shock expansion as the
accretion rate drops quickly. This occurs later in the \code{VERTEX}
simulation.

\subsection{Validity of Transport Approximations}
\label{sec:approxs}

To test the validity of our heavy-lepton neutrino approximation we
perform two additional simulations. First we carry out a collapse
simulation of the 15-$M_\odot$ model without our approximation but using
instead a kernel-based treatment of
$e^+ + e^-\leftrightarrow \nu_x + \bar{\nu}_x $. We compare it to a
simulation using our approximate method. As a reminder, our
approximation is to assign an emissivity ($\eta^{pp}$) assuming no
final state neutrino blocking and use an effective absorption cross
section computed via $\kappa^{pp}_a = \eta^{pp}/B_\nu$, where $B_\nu$ is
the value of the black body function for that energy. This removes
coupling between energy groups and species. In both cases we only
implement these rates for heavy-lepton neutrinos. For a one-to-one
comparison, these simulations we omit nucleon-nucleon Bremsstrahlung
since \code{NuLib} does not have kernels for this process.

\begin{figure}[t] \centering
  \includegraphics[width=\columnwidth]{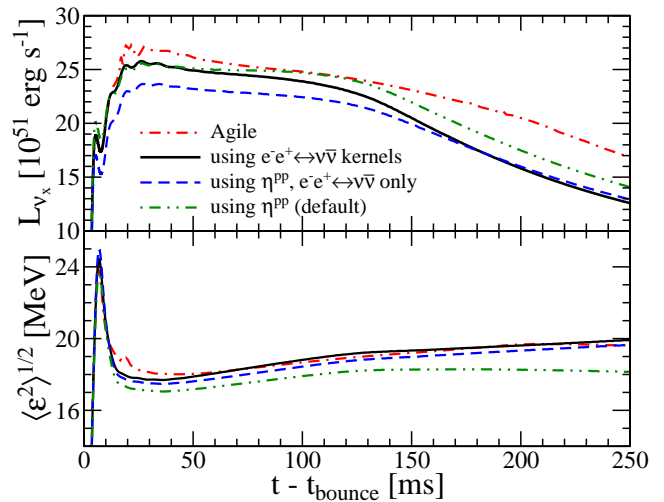}
  \caption{Investigating the pair production approximations in
    \code{GR1D}. We show the emitted heavy-lepton luminosity (top
    panel) and root mean squared energy (bottom panel) for several
    simulations designed to study the effect of our approximations.
    Our standard simulation is shown as the green dashed-dotted-dotted
    line and corresponds to the results in
    \fref{fig:lumsandenergies}. We also show results from \code{GR1D}
    using this approximation but only including
    $e^- + e^+ \leftrightarrow \nu_x + \bar{\nu}_x$ as the blue dashed
    line and the results of an implicit approach with production and
    annihilation kernels for
    $e^- + e^+ \leftrightarrow \nu_x + \bar{\nu}_x$ as the solid black
    line. For reference we show the \emph{Agile}-\code{BOLTZTRAN}
    results as a thick red dashed-dotted line. There is very little
    influence on the electron neutrino and antineutrino luminosities
    and root mean squared energies between these
    simulations.}\label{fig:heavylepton}
\end{figure}

In \fref{fig:heavylepton}, we show the results of these two
simulations (the solid line for the kernel-based treatment and the
dashed line for the approximation). For clarity, we only show the
heavy-lepton neutrino luminosities (top panel) and root mean squared
energies (bottom panel). The electron-type neutrino quantities are not
directly influenced by this treatment and change very little. In
addition, we include the \emph{Agile}-\code{BOLTZTRAN} results
(dashed-dotted line) for comparison. Our approximation results in a
$\sim$7\% lower value of the luminosity for the first $\sim$150\,ms.
Coincidentally, this $\sim 7\%$ difference is also the difference one
expects for excluding nucleon-nucleon Bremsstrahlung. The latter was
noted in \cite{lentz:12b}, but we can also see the effect in our
simulations. Also shown in \fref{fig:heavylepton} (as the
dashed-dotted-dotted line) is the reference simulation from the
previous section that uses our heavy-lepton approximation but includes
both electron-positron annihilation and nucleon-nucleon Bremsstrahlung
as sources. The difference between this line and the dashed line is
the effect of nucleon-nucleon Bremsstrahlung. These two results
together explain the similarity of the kernel-based luminosity and the
approximation seen in \fref{fig:heavylepton} and predict that a full
kernel-based treatment would see a luminosity that is roughly
$\sim$7\% higher than the kernel-based treatment with only
electron-positron annihilation. Such a luminosity would be in better
agreement with the results of \cite{liebendoerfer:05} and
\cite{mueller:10}.

The heavy-lepton root mean squared energies (bottom panel of
\fref{fig:heavylepton}) are also well captured by our pair production
approximation. When comparing the proper kernel treatment to the
equivalent simulation that uses the approximation (solid line to the
dashed line), our approximation gives a slightly ($\sim$2\%, or
0.3\,MeV) lower root mean squared energy consistently throughout the
first 250\,ms of post-bounce evolution. All of these variations in the
root mean squared energy are within the error bounds set by other
simulation codes (c.f. \citealt{liebendoerfer:05, marek:06,
mueller:10}). Inclusion of nucleon-nucleon Bremsstrahlung in our
approximate method (dashed-dotted-dotted line in the bottom panel of
\fref{fig:heavylepton}) reduces the root mean squared energy by a
further $\sim$1\,MeV.

The second approximation is our explicit treatment of inelastic
neutrino-electron scattering. The results above treat
neutrino-electron scattering explicitly throughout the entire
evolution. As mentioned in \sref{sec:nusourceterms}, we place a
suppression on the high density ($\rho >
5\times10^{12}$\,g\,cm$^{-3}$) scattering kernels to enable our
explicit treatment. This kernel suppression is necessary to keep the
scattering amplitudes small and is the main source of error, rather
than any explicit versus implicit implementation differences. To test
this, we perform a simulation where we treat inelastic
neutrino-electron scattering implicitly. We see essentially no
differences in the hydrodynamic quantities and the electron neutrino
and antineutrino luminosities and root mean squared energies. The
largest differences arise in the heavy-lepton sector, as expected,
since these neutrinos are being emitted from the highest density
regions where our artificial reduction of the kernels is being
implemented. To fully see the effect of this approximation, we test
our explicit/implicit inelastic neutrino-electron scattering
approximation using an implicit treatment of heavy-lepton neutrino
production. The observed differences in the heavy-lepton neutrino
luminosities and root mean squared energies reach the 2\% level around
200\,ms. As the heavy-lepton neutrinosphere recedes to higher
densities this difference increases. Our suppression of the scattering
kernels at high density reduces the efficiency of down-scattering
neutrinos and the emitted spectrum has a larger root mean squared
energy and a lower overall neutrino luminosity.

\subsection{Neutrino Number Conservation}

We test neutrino number conservation via a simplified scenario
described here. We take several configurations from the standard run
in \sref{sec:s15compare} and evolve the neutrino radiation field while
keeping the hydrodynamics fixed.  After sufficient evolution time has
passed, the neutrino radiation fields reach equilibrium.  At that
point, the difference between the total number of neutrinos entering the
grid (through matter interactions) and exiting the grid (through the
outer boundary) directly conveys any non-conservation in the evolution
equations.

During the late stages of core collapse, both inelastic scattering and
energy coupling terms are important and are potential sources of
number violation.  We examine two specific configurations with central
densities of $\rho_c \sim 10^{11}$\,g\,cm$^{-3}$ and $\rho_c \sim 9
\times 10^{11}$\,g\,cm$^{-3}$.  To isolate causes of neutrino number
violation, we test these configurations with and without
neutrino-electron inelastic scattering.  At $\rho_c \sim
10^{11}$\,g\,cm$^{-3}$, the neutrino number violation is
$\lesssim$0.27\% regardless of whether we include inelastic scattering
or not.  At $\rho_c \sim 9\times 10^{11}$\,g\,cm$^{-3}$, the neutrino
number violation seen is $\sim$0.76\% and $\sim$0.53\% when including
and neglecting neutrino-electron inelastic scattering, respectively.

We also do this test at 100\,ms after bounce. We find $\sim$3\% number
violation for electron neutrinos regardless of whether we include
inelastic scattering or not. This result is consistent with the tests
in section \sref{sec:gravitatingspheres} where we saw that most of the
number violation ($\sim$2\% in those tests) is occurring in the
optically thick core.

\subsection{Black Hole Formation}

\begin{figure*}[t]
\centering
\includegraphics[width=0.94\columnwidth]{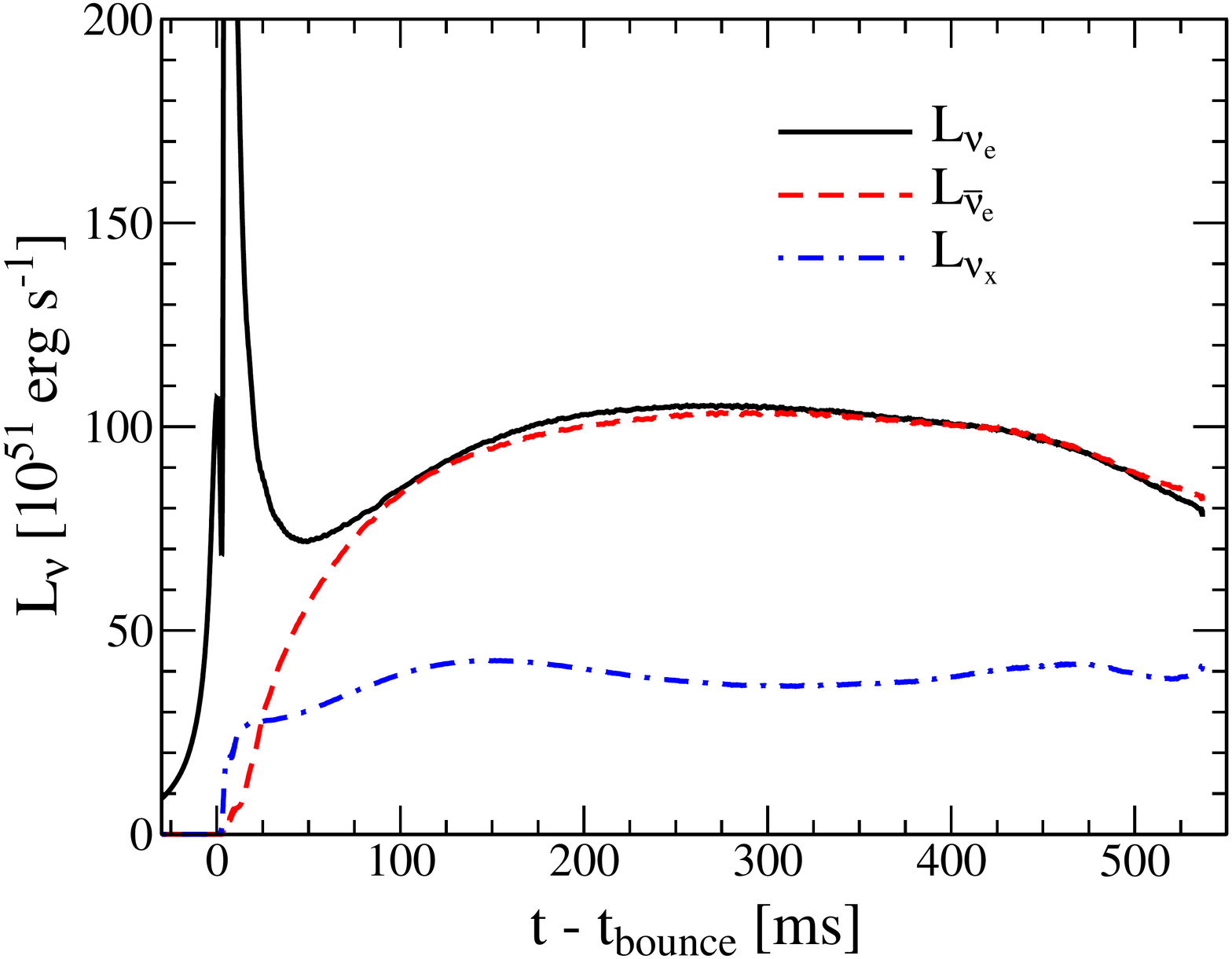}\ \ \ \ \ \
\ \ \includegraphics[width=0.94\columnwidth]{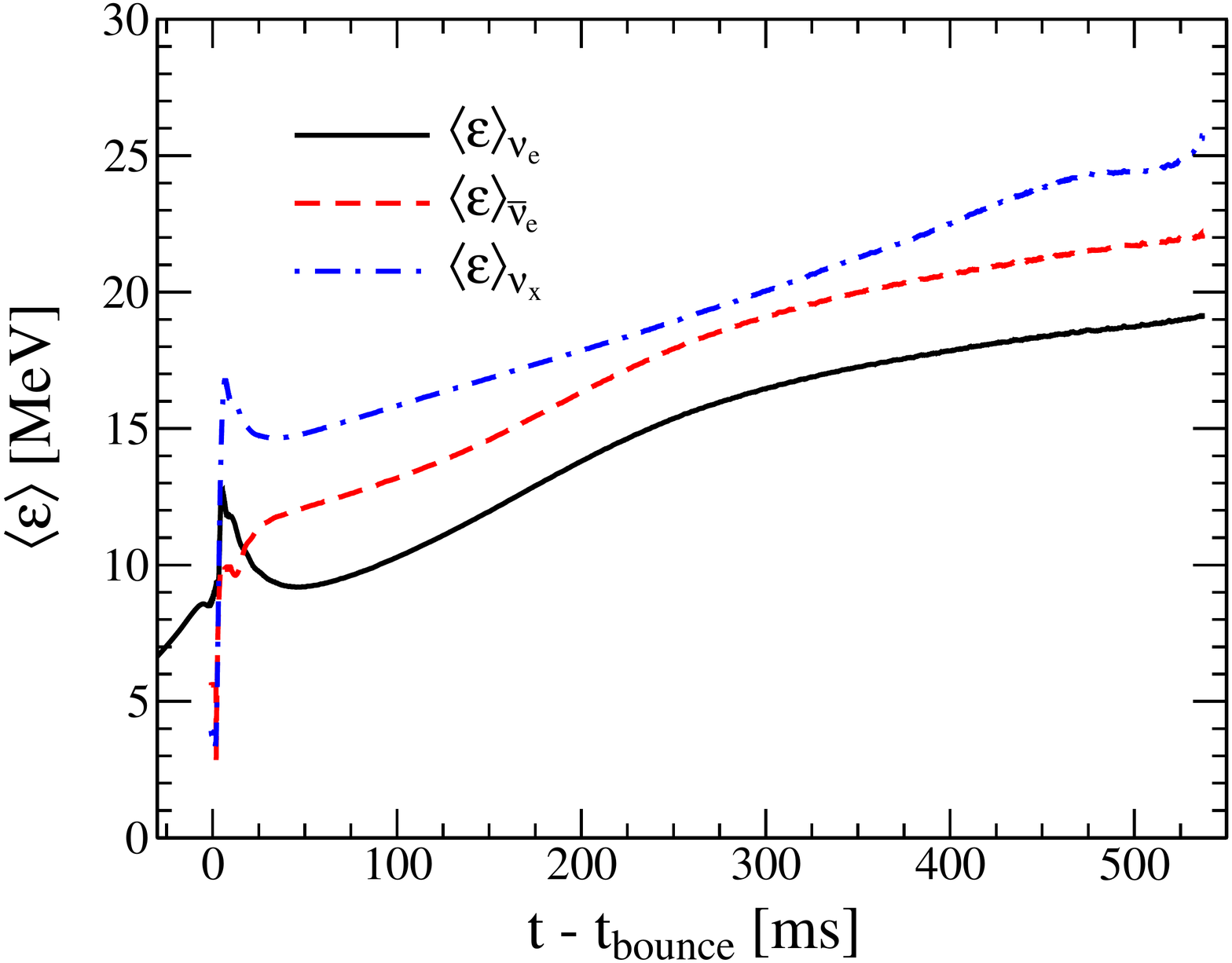} 
\caption{Neutrino observables from a failed CCSN simulation of a
  40\,$M_\odot$ progenitor star from \cite{woosley:07} evolved with
  the LS220 EOS. We show the neutrino luminosity (left panel) and the
  neutrino average energy (right panel). In both panels, the curves
  corresponding to electron neutrinos are shown as solid black lines,
  electron antineutrino curves are shown as dashed red lines, and
  heavy-lepton neutrino curves are shown as a dashed-dotted blue line.
  Note the luminosities and average energies presented here are those
  as measured in the lab frame at 500\,km. The lapse function at
  500\,km is $\alpha \sim$0.99, therefore very little additional
  redshifting will take place as the neutrinos travel to infinity.
  This is different than \fref{fig:lumsandenergies} where the
  luminosities are measured in the fluid (or comoving) frame for the
  sake of comparison. In order to compute the neutrino average energy
  in the lab frame we use the fluid frame value (where the energies
  are defined) and convert to the lab frame via
  $\langle \epsilon \rangle_\mathrm{lab} = \langle \epsilon
  \rangle_\mathrm{fluid} W (1+v)$.
  Protoneutron star collapse to a black hole occurs at $\sim$537\,ms,
  due to the finite neutrino transport time, the last $\sim$1.7\,ms of
  the neutrino signal has not yet reached the observer at
  500\,km.}\label{fig:s40lumsandenergies} \end{figure*}

As a final test of our general relativistic neutrino transport code,
we evolve a presupernova model with a zero-age main sequence mass of 40\,$M_\odot$
to the onset of protoneutron star collapse to a black hole. We use the
progenitor model from \cite{woosley:07} and the $K_0=220\,$MeV variant
of the Lattimer \& Swesty EOS (LS220) and include the weak magnetism
and recoil corrections in our neutrino interaction rates. \code{GR1D}
evolves to the onset of protoneutron star collapse impressively well
considering the strong gravity and relativistic fluid flow. The largest issue we
face occurs when the shock recedes to very small radii and the
pre-shock velocity (and its gradient) becomes very large. For example,
at 100\,ms pre-black hole formation the fluid velocity outside of the
shock is 0.35$c$, the $g_{rr}$ component of the metric reaches a
maximum of 1.7, and the central lapse is 0.5. The explicit energy
coupling procedure we have used up to now begins to fail at the shock
front.  This is understandable as, for example, the average energy of
neutrinos in the fluid frame changes by $\sim 40\%$ across a few zones
at the shock front. To aid \code{GR1D} in finding the transport
solution we switch to an implicit treatment of the energy coupling
roughly 100\,ms before black hole formation. We also find it necessary
to reduce the Courant factor when the simulation approaches black hole
formation. The metric of \code{GR1D} does not allow the existence of
an apparent horizon, and therefore we are unable to evolve past black
hole formation, the code stops when the central lapse is $\sim 0.023$
and the central density has reached
$\sim 3.4\times10^{15}$\,g\,cm$^{-3}$. We show the evolution of the
neutrino luminosity (left panel) and average energy (right panel) in
\fref{fig:s40lumsandenergies} from this black hole formation
simulation. Unlike in the previous section, here we show the neutrino
quantities in the lab frame. The lab frame values are much closer to
the asymptotic value than the fluid frame values as the fluid velocity
at 500\,km is $\sim 0.1\,c$. Since the neutrino radiation takes at
least $t\,=$\,500\,km/c$\,\sim 1.7$\,ms to travel from the
protoneutron star to the boundary at 500\,km, we do not know what the
neutrino signal from the final $\sim$\,ms of evolution is. The
post-bounce neutrino luminosities are much higher than the
$15\,M_\odot$ model explored in the last section. This is because the
post-bounce accretion rate is substantially higher in this
progenitor. The qualitative features of the neutrino luminosity and
the root mean squared energy match those of \cite{fischer:09a} who
used \emph{Agile}-\code{BOLTZTRAN} to study black hole formation in a
different progenitor (a $40\,M_\odot$ progenitor from \citealt{ww:95})
and with a different EOS (LS180, and the H. Shen EOS). The electron
neutrino and electron antineutrino luminosities peak at $\sim 300\,$ms
after bounce. After this they slowly decline until $\sim 450\,$ms
after bounce when the decline steepens. Similar to the 15\,$M_\odot$
model from the previous section, this sharp decline is due to the
accretion of the silicon-oxygen interface which is located at a
baryonic mass coordinate of 2.34$\,M_\odot$ and accretes through the
shock at this time. Gravitational time dilation and redshift also
contribute to this slow decline over the last $\sim 200$\,ms. The
evolution of the heavy-lepton neutrino luminosity is more complex
since there are several effects coming together. Unlike electron
neutrinos and antineutrinos, heavy-lepton neutrino production is not
predominately fueled by accretion. It comes from deeper in the
protoneutron star and is more analogous to a cooling luminosity. It
initially plateaus much earlier ($\sim 150$\,ms) than the electron
type luminosities. Normally, protoneutron star contraction is
regulated by cooling and the luminosity slowly declines over time.
However, in the case of these extremely massive protoneutron stars
with high post-bounce accretion rates, the heavy-lepton neutrino
luminosity is driven back up at late times as the matter emitting
these neutrinos is not able to cool fast enough via neutrino emission
to counteract the compressional heating from gravity and accretion.
The same heavy-lepton neutrino luminosity structure is seen in
\cite{fischer:09a}. The final bump in the heavy-lepton neutrino
luminosity is a combination of the drop due to the silicon-oxygen
interface accreting in and the rise due to the protoneutron star
contraction. As the protoneutron star becomes progressively more
massive and compact, the neutrino energies also increase.

With \code{GR1D}'s neutrino leakage scheme we found a black hole
formation time of 561\,ms and a maximum protoneutron star
gravitational (baryonic) mass of $\sim2.31\,M_\odot$
($\sim$2.44\,$M_\odot$) \citep{oconnor:11}. With our neutrino
transport methods we find a black hole formation time of $\sim$537\,ms
($\sim$24\,ms before the leakage calculation) and a maximum
protoneutron star gravitational (baryonic) mass of
$\sim$2.251\,$M_\odot$ ($\sim$2.377\,$M_\odot$). These results are
remarkably close and confirm our previous work that the progenitor
structure, and not details of the neutrino physics, is the determining
factor in black hole formation properties \citep{oconnor:11}. Our
leakage scheme was unable to reliably predict the total neutrino
emission. However, with our transport scheme we can make a reliable
prediction on the total energy and neutrino number emitted from this
particular failed supernova (i.e.\ for a progenitor matching the
40\,$M_\odot$ star from \cite{woosley:07} with the LS220 EOS). We find
a total neutrino number emission of $\sim 2.56\times 10^{57}$,
$\sim 2.33\times 10^{57}$, and $\sim 4.03\times10^{57}$, for electron
neutrino, electron antineutrino, and all four heavy-lepton neutrinos,
respectively. The total energy emission is
$\sim 54.4\times 10^{51}$\,erg, $\sim 47.6\times 10^{51}$\,erg, and
$\sim 80.6\times 10^{51}$\,erg for electron neutrino, electron
antineutrino, and all four heavy-lepton neutrinos, respectively.
Summed, this corresponds to $\sim 182.6\times 10^{51}$\,erg or
equivalently $\sim 0.102\,M_\odot$ of mass. The remaining difference
between the gravitational mass and the baryonic mass
($\sim 0.02\,M_\odot$) was present in the initial progenitor model. We
note that while this simulation corresponds to a failed supernova, it
only radiates $\sim$50\% of the energy expected to be radiated in
successful CCSNe. The rest of the binding energy released during the
collapse is still trapped in the matter (either as thermal energy or
trapped neutrinos) at the point when the protoneutron star begins its
collapse.

\section{Conclusions}
\label{sec:conclusions}

Neutrinos play a crucial, if not dominant, role in reviving the
stalled accretion shock that forms after the iron-core collapse of an
evolved massive star. In order to achieve an accurate and
self-consistent treatment of neutrinos in core collapse simulations
one has to consider several important aspects of the problem. Deep in
the protoneutron star, the mean free path of neutrinos is very small.
However, by the time the neutrinos reach 50-130\,km, the opacity has
decreased enough so that the neutrinos are essentially decoupled from
the matter and are free streaming. This transition region is between
the optically thick and optically thin region and is very important to
capture correctly since it is where the net neutrino heating takes
place. Another critical aspect of the problem that must be considered
is the strong energy dependence of the neutrino interaction rates.
This leads to neutrinos of different energies decoupling at different
densities and radii and therefore any self-consistent treatment must
be done in an energy dependent way.

For the \emph{hydrodynamic} evolution in the CCSN problem we do not
have to deal with these issues because the matter particles are always
in thermodynamic equilibrium. We can completely ignore the momentum
dependence of the particles (other than the net value) and just solve
the hydrodynamic conservation laws for mass, energy, momentum in one,
two, or three spatial dimensions (plus time). Since neutrinos in CCSNe
are not always in thermodynamic equilibrium, we cannot apply the same
techniques for neutrino transport. This makes the symmetry free
problem not three dimensional (plus time) but rather a six dimensional
problem (plus time). Simulating this six dimensional system at the
resolution we need to capture all the essential physics of the CCSN
central engine is not feasible with current computational power, so
some approximations must still be made. In this paper, we reduced the
dimensionality of the problem by removing the angular dependence from
the neutrino distribution function and instead evolved moments of the
neutrino distribution function--the total energy, and the total
momentum. In this sense, our approximation is very much like the
approximation made to derive the hydrodynamic equations. The
equivalent to the matter pressure is the Eddington tensor. We applied
an analytic closure in order to derive this Eddington tensor. We
retained the energy dependence of the neutrino distribution function.
This reduces the symmetry free problem to four dimensions (plus time)
and the spherically symmetric problem to two (plus time).

The general-relativistic methodology of this neutrino transport method
has recently been presented in \cite{shibata:11} and
\cite{cardall:12rad}. We presented in this paper the spherically
symmetric equivalent in the metric and notation of \code{GR1D}. Rather
than focus on the derivation of the transport method itself (which is
more than adequately presented in both \citealt{shibata:11} and
\cite{cardall:12rad}), we focused on the numerical implementation in
spherical symmetry with the aim of developing methods that will carry
over well to multiple dimensions and parallel environments. We
presented simple radiation test cases to show our code is correctly
modeling transport phenomena in the general relativistic regime. We
also showed that our explicit treatment of the spatial flux works well
in the diffusion limit, which permitted us to forgo large and
computationally expensive matrix inversions. This will be a crucial
advantage for multidimensional simulations. We also performed a CCSN
simulation following the established standard of
\cite{liebendoerfer:05,marek:06,mueller:10}. The excellent agreement
across many quantities assures us that the neutrino transport, and its
coupling to the hydrodynamics is correctly implemented. We tested
several approximations that will make the transition from spherically
symmetric simulations to multidimensional simulations easier by
removing the need to invert large matrices. The accuracy cost of these
approximations is very little and within the differences seen in
various codes.

As a final test, we followed the evolution of a $40\,M_\odot$, solar
metallicity progenitor from the onset of core collapse, through bounce
to black hole formation. This was an especially stringent test of the
robustness of the transport scheme as the spacetime curvature becomes
strong near the end of the simulation. Furthermore, the pre-shock
velocities reached in the late stages approached $0.5\,c$, much larger
than typical CCSNe.  The black hole formation time and the black hole
birth mass agreed closely with results using a neutrino leakage
scheme. While no direct comparison can be made, the evolution of the
neutrino quantities qualitatively agrees with other neutrino transport
code studying similar progenitors with similar EOS.

In the interest of open-science, ensuring reproducibility, full
disclosure, and to provide technology for other scientific
researchers, all of our code is open-source. The neutrino transport
methods are included as an update to \code{GR1D} and are available as
a git repository at \url{www.GR1Dcode.org}. All of the neutrino
microphysics comes from the open-source neutrino interaction library
\code{NuLib}, also available as a git repository
\url{http://www.nulib.org}. Both repositories have a tagged released
named `GR1Dv2'. We make the NuLib tables used for our core collapse
simulations in this paper as well as the parameter files and scripts
needed to generate the data in this paper available at
\url{www.stellarcollapse.org/GR1Dv2}.

\section*{Acknowledgements}

The author would like to thank Francois Foucart and Luke Roberts for
many in depth discussions surrounding this topic, Matthias
Liebend\"orfer for discussions regarding \cite{liebendoerfer:05}, and
the referee for helpful comments that have improved the
manuscript. Furthermore, for productive and helpful discussions we
acknowledge Adam Burrows, Christian Cardall, Sean Couch, Thomas Janka,
Anthony Mezzacappa, Bernhard Mueller, and Chris Sullivan. The author
would like to especially thank Christian Ott for his advisement during
the first part of this work and for a careful reading of this
manuscript.  Computations were performed on the Zwicky cluster at
Caltech, which is supported by the Sherman Fairchild Foundation and by
NSF award PHY-0960291. Part of this research performed at Caltech was
supported by AST-1212170 and AST-0855535. Support for this work was
provided by NASA through Hubble Fellowship grant \#51344.001-A awarded
by the Space Telescope Science Institute, which is operated by the
Association of Universities for Research in Astronomy, Inc., for NASA,
under contract NAS 5-26555.


\begin{thebibliography}{81}
\expandafter\ifx\csname natexlab\endcsname\relax\def\natexlab#1{#1}\fi

\bibitem[{{Abdikamalov} {et~al.}(2012){Abdikamalov}, {Burrows}, {Ott},
  {L{\"o}ffler}, {O'Connor}, {Dolence}, \& {Schnetter}}]{abdikamalov:12}
{Abdikamalov}, E., {Burrows}, A., {Ott}, C.~D., {L{\"o}ffler}, F., {O'Connor},
  E., {Dolence}, J.~C., \& {Schnetter}, E. 2012, \apj, 755, 111

\bibitem[{{Antoniadis} {et~al.}(2013){Antoniadis}, {Freire}, {Wex}, {Tauris},
  {Lynch}, {van Kerkwijk}, {Kramer}, {Bassa}, {Dhillon}, {Driebe}, {Hessels},
  {Kaspi}, {Kondratiev}, {Langer}, {Marsh}, {McLaughlin}, {Pennucci}, {Ransom},
  {Stairs}, {van Leeuwen}, {Verbiest}, \& {Whelan}}]{antoniadis:13}
{Antoniadis}, J., {Freire}, P.~C.~C., {Wex}, N., {Tauris}, T.~M., {Lynch},
  R.~S., {van Kerkwijk}, M.~H., {Kramer}, M., {Bassa}, C., {Dhillon}, V.~S.,
  {Driebe}, T., {Hessels}, J.~W.~T., {Kaspi}, V.~M., {Kondratiev}, V.~I.,
  {Langer}, N., {Marsh}, T.~R., {McLaughlin}, M.~A., {Pennucci}, T.~T.,
  {Ransom}, S.~M., {Stairs}, I.~H., {van Leeuwen}, J., {Verbiest}, J.~P.~W., \&
  {Whelan}, D.~G. 2013, Science, 340, 448

\bibitem[{{Audit} {et~al.}(2002){Audit}, {Charrier}, {Chi{\`e}ze}, \&
  {Dubroca}}]{audit:02}
{Audit}, E., {Charrier}, P., {Chi{\`e}ze}, J., \& {Dubroca}, B. 2002,
  arXiv:0206281

\bibitem[{{Bethe}(1990)}]{bethe:90}
{Bethe}, H.~A. 1990, Rev. Mod. Phys., 62, 801

\bibitem[{{Bethe} \& {Wilson}(1985)}]{bethewilson:85}
{Bethe}, H.~A., \& {Wilson}, J.~R. 1985, \apj, 295, 14

\bibitem[{{Brandt} {et~al.}(2011){Brandt}, {Burrows}, {Ott}, \&
  {Livne}}]{brandt:11}
{Brandt}, T.~D., {Burrows}, A., {Ott}, C.~D., \& {Livne}, E. 2011, \apj, 728, 8

\bibitem[{{Bruenn}(1985)}]{bruenn:85}
{Bruenn}, S.~W. 1985, \apjs, 58, 771

\bibitem[{{Buras} {et~al.}(2006{\natexlab{a}}){Buras}, {Janka}, {Rampp}, \&
  {Kifonidis}}]{buras:06b}
{Buras}, R., {Janka}, H.-T., {Rampp}, M., \& {Kifonidis}, K.
  2006{\natexlab{a}}, \aap, 457, 281

\bibitem[{{Buras} {et~al.}(2006{\natexlab{b}}){Buras}, {Rampp}, {Janka}, \&
  {Kifonidis}}]{buras:06a}
{Buras}, R., {Rampp}, M., {Janka}, H.-T., \& {Kifonidis}, K.
  2006{\natexlab{b}}, \aap, 447, 1049

\bibitem[{{Burrows}(2013)}]{burrows:13a}
{Burrows}, A. 2013, Rev. Mod. Phys., 85, 245

\bibitem[{{Burrows} {et~al.}(1995){Burrows}, {Hayes}, \& {Fryxell}}]{bhf:95}
{Burrows}, A., {Hayes}, J., \& {Fryxell}, B.~A. 1995, \apj, 450, 830

\bibitem[{{Burrows} {et~al.}(2007){Burrows}, {Livne}, {Dessart}, {Ott}, \&
  {Murphy}}]{burrows:07a}
{Burrows}, A., {Livne}, E., {Dessart}, L., {Ott}, C.~D., \& {Murphy}, J. 2007,
  Astrophys. J., 655, 416

\bibitem[{{Burrows} {et~al.}(2006){Burrows}, {Reddy}, \& {Thompson}}]{brt:06}
{Burrows}, A., {Reddy}, S., \& {Thompson}, T.~A. 2006, Nuclear Physics A, 777,
  356

\bibitem[{{Burrows} {et~al.}(2000){Burrows}, {Young}, {Pinto}, {Eastman}, \&
  {Thompson}}]{burrows:00}
{Burrows}, A., {Young}, T., {Pinto}, P., {Eastman}, R., \& {Thompson}, T.~A.
  2000, \apj, 539, 865

\bibitem[{{Cardall} {et~al.}(2013){Cardall}, {Endeve}, \&
  {Mezzacappa}}]{cardall:12rad}
{Cardall}, C.~Y., {Endeve}, E., \& {Mezzacappa}, A. 2013, \prd, 87, 103004

\bibitem[{{Chernohorsky}(1994)}]{cernohorsky:94b}
{Chernohorsky}, J. 1994, \apj, 433, 247

\bibitem[{{Colella} \& {Woodward}(1984)}]{colella:84}
{Colella}, P., \& {Woodward}, P.~R. 1984, J. Comp. Phys., 54, 174

\bibitem[{{Demorest} {et~al.}(2010){Demorest}, {Pennucci}, {Ransom}, {Roberts},
  \& {Hessels}}]{demorest:10}
{Demorest}, P.~B., {Pennucci}, T., {Ransom}, S.~M., {Roberts}, M.~S.~E., \&
  {Hessels}, J.~W.~T. 2010, \nat, 467, 1081

\bibitem[{{Einfeldt}(1988)}]{HLLE:88}
{Einfeldt}, B. 1988, in Shock tubes and waves; Proceedings of the Sixteenth
  International Symposium, Aachen, Germany, July 26--31, 1987. VCH Verlag,
  Weinheim, Germany, 671

\bibitem[{{Fischer} {et~al.}(2009){Fischer}, {Whitehouse}, {Mezzacappa},
  {Thielemann}, \& {Liebend{\"o}rfer}}]{fischer:09a}
{Fischer}, T., {Whitehouse}, S.~C., {Mezzacappa}, A., {Thielemann}, F.-K., \&
  {Liebend{\"o}rfer}, M. 2009, \aap, 499, 1

\bibitem[{{Fischer} {et~al.}(2010){Fischer}, {Whitehouse}, {Mezzacappa},
  {Thielemann}, \& {Liebend{\"o}rfer}}]{fischer:10}
---. 2010, \aap, 517, A80

\bibitem[{{Font} {et~al.}(2000){Font}, {Miller}, {Suen}, \& {Tobias}}]{font:00}
{Font}, J.~A., {Miller}, M., {Suen}, W.-M., \& {Tobias}, M. 2000, \prd, 61,
  044011

\bibitem[{{Fryer}(1999)}]{fryer:99}
{Fryer}, C.~L. 1999, \apj, 522, 413

\bibitem[{{Gourgoulhon}(1991)}]{gourgoulhon:91}
{Gourgoulhon}, E. 1991, \aap, 252, 651

\bibitem[{{Herant} {et~al.}(1994){Herant}, {Benz}, {Hix}, {Fryer}, \&
  {Colgate}}]{herant:94}
{Herant}, M., {Benz}, W., {Hix}, W.~R., {Fryer}, C.~L., \& {Colgate}, S.~A.
  1994, \apj, 435, 339

\bibitem[{{Horowitz}(1997)}]{horowitz:97}
{Horowitz}, C.~J. 1997, \prd, 55, 4577

\bibitem[{{Horowitz}(2002)}]{horowitz:02}
---. 2002, \prd, 65, 043001

\bibitem[{{H{\"u}depohl} {et~al.}(2010){H{\"u}depohl}, {M{\"u}ller}, {Janka},
  {Marek}, \& {Raffelt}}]{huedepohl:10}
{H{\"u}depohl}, L., {M{\"u}ller}, B., {Janka}, H.-T., {Marek}, A., \&
  {Raffelt}, G.~G. 2010, Phys. Rev. Lett., 104, 251101

\bibitem[{{Janka}(1992)}]{janka:92c}
{Janka}, H.-T. 1992, \aap, 256, 452

\bibitem[{{Janka}(2012)}]{janka:12a}
---. 2012, Ann. Rev. Nuc. Par. Sci., 62, 407

\bibitem[{{Janka} \& {Hillebrandt}(1989{\natexlab{a}})}]{janka:89c}
{Janka}, H.-T., \& {Hillebrandt}, W. 1989{\natexlab{a}}, \aaps, 78, 375

\bibitem[{{Janka} \& {Hillebrandt}(1989{\natexlab{b}})}]{janka:89d}
---. 1989{\natexlab{b}}, \aap, 224, 49

\bibitem[{{Janka} {et~al.}(2007){Janka}, {Langanke}, {Marek},
  {Mart{\'{\i}}nez-Pinedo}, \& {M{\"u}ller}}]{janka:07}
{Janka}, H.-T., {Langanke}, K., {Marek}, A., {Mart{\'{\i}}nez-Pinedo}, G., \&
  {M{\"u}ller}, B. 2007, \physrep, 442, 38

\bibitem[{{Jin} \& {Levermore}(1996)}]{jin:96}
{Jin}, S., \& {Levermore}, C.~D. 1996, J. Comp. Phys., 126, 449

\bibitem[{{Just} {et~al.}(2015){Just}, {Obergaulinger}, \& {Janka}}]{just:15}
{Just}, O., {Obergaulinger}, M., \& {Janka}, H.-T. 2015, ArXiv e-prints
  arXiv:1501.02999

\bibitem[{{Kuroda} {et~al.}(2012){Kuroda}, {Kotake}, \& {Takiwaki}}]{kuroda:12}
{Kuroda}, T., {Kotake}, K., \& {Takiwaki}, T. 2012, \apj, 755, 11

\bibitem[{{Kuroda} {et~al.}(2015){Kuroda}, {Takiwaki}, \& {Kotake}}]{kuroda:15}
{Kuroda}, T., {Takiwaki}, T., \& {Kotake}, K. 2015, ArXiv e-prints

\bibitem[{{Langanke} {et~al.}(2003){Langanke}, {Mart{\'{\i}}nez-Pinedo},
  {Sampaio}, {Dean}, {Hix}, {Messer}, {Mezzacappa}, {Liebend{\"o}rfer},
  {Janka}, \& {Rampp}}]{langanke:03}
{Langanke}, K., {Mart{\'{\i}}nez-Pinedo}, G., {Sampaio}, J.~M., {Dean}, D.~J.,
  {Hix}, W.~R., {Messer}, O.~E., {Mezzacappa}, A., {Liebend{\"o}rfer}, M.,
  {Janka}, H.-T., \& {Rampp}, M. 2003, \prl, 90, 241102

\bibitem[{Lattimer \& Swesty(1991)}]{lseos:91}
Lattimer, J.~M., \& Swesty, F.~D. 1991, {Nucl. Phys. A}, 535, 331

\bibitem[{{Lentz} {et~al.}(2012){Lentz}, {Mezzacappa}, {Bronson Messer}, {Hix},
  \& {Bruenn}}]{lentz:12b}
{Lentz}, E.~J., {Mezzacappa}, A., {Bronson Messer}, O.~E., {Hix}, W.~R., \&
  {Bruenn}, S.~W. 2012, \apj, 760, 94

\bibitem[{Liebend\"orfer(2015)}]{liebendoerfer:15pc}
Liebend\"orfer, M. 2015, Private communiciation

\bibitem[{{Liebend{\"o}rfer} {et~al.}(2004){Liebend{\"o}rfer}, {Messer},
  {Mezzacappa}, {Bruenn}, {Cardall}, \& {Thielemann}}]{liebendoerfer:04}
{Liebend{\"o}rfer}, M., {Messer}, O.~E.~B., {Mezzacappa}, A., {Bruenn}, S.~W.,
  {Cardall}, C.~Y., \& {Thielemann}, F.-K. 2004, \apjs, 150, 263

\bibitem[{{Liebend{\"o}rfer} {et~al.}(2005){Liebend{\"o}rfer}, {Rampp},
  {Janka}, \& {Mezzacappa}}]{liebendoerfer:05}
{Liebend{\"o}rfer}, M., {Rampp}, M., {Janka}, H.-T., \& {Mezzacappa}, A. 2005,
  \apj, 620, 840

\bibitem[{{Liebend{\"o}rfer} {et~al.}(2009){Liebend{\"o}rfer}, {Whitehouse}, \&
  {Fischer}}]{liebendoerfer:09}
{Liebend{\"o}rfer}, M., {Whitehouse}, S.~C., \& {Fischer}, T. 2009, \apj, 698,
  1174

\bibitem[{{Lindquist}(1966)}]{lindquist:66}
{Lindquist}, R.~W. 1966, Annals of Physics, 37, 487

\bibitem[{{Livne} {et~al.}(2004){Livne}, {Burrows}, {Walder}, {Lichtenstadt},
  \& {Thompson}}]{livne:04}
{Livne}, E., {Burrows}, A., {Walder}, R., {Lichtenstadt}, I., \& {Thompson},
  T.~A. 2004, \apj, 609, 277

\bibitem[{{Marek} {et~al.}(2006){Marek}, {Dimmelmeier}, {Janka}, {M{\"u}ller},
  \& {Buras}}]{marek:06}
{Marek}, A., {Dimmelmeier}, H., {Janka}, H.-T., {M{\"u}ller}, E., \& {Buras},
  R. 2006, \aap, 445, 273

\bibitem[{{Mezzacappa} \& {Bruenn}(1993{\natexlab{a}})}]{mezzacappa:93b}
{Mezzacappa}, A., \& {Bruenn}, S.~W. 1993{\natexlab{a}}, \apj, 405, 669

\bibitem[{{Mezzacappa} \& {Bruenn}(1993{\natexlab{b}})}]{mezzacappa:93c}
---. 1993{\natexlab{b}}, \apj, 410, 740

\bibitem[{{Mezzacappa} \& {Bruenn}(1993{\natexlab{c}})}]{mezzacappa:93a}
---. 1993{\natexlab{c}}, \apj, 405, 637

\bibitem[{{Minerbo}(1978)}]{minerbo:78}
{Minerbo}, G.~N. 1978, \jqsrt, 20, 541

\bibitem[{{M{\"u}ller} \& {Janka}(2015)}]{mueller:15}
{M{\"u}ller}, B., \& {Janka}, H.-T. 2015, \mnras, 448, 2141

\bibitem[{{M{\"u}ller} {et~al.}(2010){M{\"u}ller}, {Janka}, \&
  {Dimmelmeier}}]{mueller:10}
{M{\"u}ller}, B., {Janka}, H.-T., \& {Dimmelmeier}, H. 2010, \apjs, 189, 104

\bibitem[{{Obergaulinger} {et~al.}(2014){Obergaulinger}, {Janka}, \&
  {Aloy}}]{obergaulinger:14}
{Obergaulinger}, M., {Janka}, H.-T., \& {Aloy}, M.~A. 2014, \mnras, 445, 3169

\bibitem[{{O'Connor} \& {Ott}(2010)}]{oconnor:10}
{O'Connor}, E., \& {Ott}, C.~D. 2010, Class. Quantum Grav., 27, 114103

\bibitem[{{O'Connor} \& {Ott}(2011)}]{oconnor:11}
---. 2011, \apj, 730, 70

\bibitem[{{O'Connor} \& {Ott}(2013)}]{oconnor:13}
---. 2013, \apj, 762, 126

\bibitem[{{Ott} {et~al.}(2008){Ott}, {Burrows}, {Dessart}, \& {Livne}}]{ott:08}
{Ott}, C.~D., {Burrows}, A., {Dessart}, L., \& {Livne}, E. 2008, \apj, 685,
  1069

\bibitem[{{Perego} {et~al.}(2014){Perego}, {Rosswog}, {Cabez{\'o}n},
  {Korobkin}, {K{\"a}ppeli}, {Arcones}, \& {Liebend{\"o}rfer}}]{perego:14}
{Perego}, A., {Rosswog}, S., {Cabez{\'o}n}, R.~M., {Korobkin}, O.,
  {K{\"a}ppeli}, R., {Arcones}, A., \& {Liebend{\"o}rfer}, M. 2014, \mnras,
  443, 3134

\bibitem[{{Pons} {et~al.}(2000){Pons}, {Ib{\'a}{\~n}ez}, \&
  {Miralles}}]{pons:00}
{Pons}, J.~A., {Ib{\'a}{\~n}ez}, J.~M., \& {Miralles}, J.~A. 2000, \mnras, 317,
  550

\bibitem[{{Rampp} \& {Janka}(2002)}]{rampp:02}
{Rampp}, M., \& {Janka}, H.-T. 2002, \aap, 396, 361

\bibitem[{{Reddy} {et~al.}(1998){Reddy}, {Prakash}, \& {Lattimer}}]{reddy:98}
{Reddy}, S., {Prakash}, M., \& {Lattimer}, J.~M. 1998, \prd, 58, 013009

\bibitem[{{Roberts}(2014)}]{roberts:14priv}
{Roberts}, L. 2014, private communication

\bibitem[{{Roberts}(2012)}]{roberts:12b}
{Roberts}, L.~F. 2012, \apj, 755, 126

\bibitem[{{Romero} {et~al.}(1996){Romero}, {Ibanez}, {Marti}, \&
  {Miralles}}]{romero:96}
{Romero}, J.~V., {Ibanez}, J.~M., {Marti}, J.~M., \& {Miralles}, J.~A. 1996,
  \apj, 462, 839

\bibitem[{{Rosswog} \& {Liebend{\"o}rfer}(2003)}]{rosswog:03b}
{Rosswog}, S., \& {Liebend{\"o}rfer}, M. 2003, \mnras, 342, 673

\bibitem[{{Ruffert} {et~al.}(1996){Ruffert}, {Janka}, \&
  {Sch\"afer}}]{ruffert:96}
{Ruffert}, M., {Janka}, H.-T., \& {Sch\"afer}, G. 1996, \aap, 311, 532

\bibitem[{{Scheck} {et~al.}(2006){Scheck}, {Kifonidis}, {Janka}, \&
  {M{\"u}ller}}]{scheck:06}
{Scheck}, L., {Kifonidis}, K., {Janka}, H.-T., \& {M{\"u}ller}, E. 2006, \aap,
  457, 963

\bibitem[{{Sekiguchi}(2010)}]{sekiguchi:10}
{Sekiguchi}, Y. 2010, \cqg, 27, 114107

\bibitem[{{Shibata} {et~al.}(2011){Shibata}, {Kiuchi}, {Sekiguchi}, \&
  {Suwa}}]{shibata:11}
{Shibata}, M., {Kiuchi}, K., {Sekiguchi}, Y., \& {Suwa}, Y. 2011, Prog. Theor.
  Phys., 125, 1255

\bibitem[{{Smit} {et~al.}(1997){Smit}, {Cernohorsky}, \& {Dullemond}}]{smit:97}
{Smit}, J.~M., {Cernohorsky}, J., \& {Dullemond}, C.~P. 1997, \aap, 325, 203

\bibitem[{{Sumiyoshi} {et~al.}(2015){Sumiyoshi}, {Takiwaki}, {Matsufuru}, \&
  {Yamada}}]{sumiyoshi:15}
{Sumiyoshi}, K., {Takiwaki}, T., {Matsufuru}, H., \& {Yamada}, S. 2015, \apjs,
  216, 5

\bibitem[{{Sumiyoshi} \& {Yamada}(2012)}]{sumiyoshi:12}
{Sumiyoshi}, K., \& {Yamada}, S. 2012, \apjs, 199, 17

\bibitem[{{Sumiyoshi} {et~al.}(2005){Sumiyoshi}, {Yamada}, {Suzuki}, {Shen},
  {Chiba}, \& {Toki}}]{sumiyoshi:05}
{Sumiyoshi}, K., {Yamada}, S., {Suzuki}, H., {Shen}, H., {Chiba}, S., \&
  {Toki}, H. 2005, \apj, 629, 922

\bibitem[{{Swesty} \& {Myra}(2009)}]{swesty:09}
{Swesty}, F.~D., \& {Myra}, E.~S. 2009, \apjs, 181, 1

\bibitem[{{Thompson} {et~al.}(2003){Thompson}, {Burrows}, \&
  {Pinto}}]{thompson:03}
{Thompson}, T.~A., {Burrows}, A., \& {Pinto}, P.~A. 2003, \apj, 592, 434

\bibitem[{van Leer(1977)}]{vanleer:77}
van Leer, B.~J. 1977, J. Comp. Phys., 23, 276

\bibitem[{{Woosley} \& {Heger}(2007)}]{woosley:07}
{Woosley}, S.~E., \& {Heger}, A. 2007, \physrep, 442, 269

\bibitem[{{Woosley} \& {Weaver}(1995)}]{ww:95}
{Woosley}, S.~E., \& {Weaver}, T.~A. 1995, \apjs, 101, 181

\bibitem[{{Yakunin} {et~al.}(2010){Yakunin}, {Marronetti}, {Mezzacappa},
  {Bruenn}, {Lee}, {Chertkow}, {Hix}, {Blondin}, {Lentz}, {Bronson Messer}, \&
  {Yoshida}}]{yakunin:10}
{Yakunin}, K.~N., {Marronetti}, P., {Mezzacappa}, A., {Bruenn}, S.~W., {Lee},
  C.-T., {Chertkow}, M.~A., {Hix}, W.~R., {Blondin}, J.~M., {Lentz}, E.~J.,
  {Bronson Messer}, O.~E., \& {Yoshida}, S. 2010, Class. Quantum Grav., 27,
  194005

\bibitem[{{Yamada}(1997)}]{yamada:97}
{Yamada}, S. 1997, \apj, 475, 720

\bibitem[{{Yueh} \& {Buchler}(1976)}]{yueh:76}
{Yueh}, W.~R., \& {Buchler}, J.~R. 1976, \apss, 41, 221

\bibitem[{{Zhang} {et~al.}(2013){Zhang}, {Howell}, {Almgren}, {Burrows},
  {Dolence}, \& {Bell}}]{zhang:13}
{Zhang}, W., {Howell}, L., {Almgren}, A., {Burrows}, A., {Dolence}, J., \&
  {Bell}, J. 2013, \apjs, 204, 7

\end{thebibliography}
\end{document}